\documentclass[prd,showpacs,superscriptaddress,preprintnumbers,twocolumn,amsmath,footinbib]{revtex4}
 \usepackage[dvips,final]{graphicx}
  \usepackage{amssymb,amsmath,epsfig,bm,pifont}
\usepackage{color}
\definecolor{darkgreen}{rgb}{0,0.66,0}
 
  \newcommand{\BluTn}[1]{\textcolor{blue}{#1}}
   \newcommand{\RedTn}[1]{\textcolor{red}{#1}}
\usepackage{relsize}
\newcommand{\babar}{{\mbox{\slshape B\kern-0.1em{\smaller A}\kern-0.1em
            B\kern-0.1em{\smaller A\kern-0.2em R}}}
           }
\newcommand{\babaR}{{\mbox{\slshape B\kern-0.1em{\smaller A}\kern-0.1em
            B\kern-0.1em{\smaller A\kern-0.2em R,}}}
           }
\newcommand{\qq}[1]{\langle\bar{q}{#1}q\rangle}          %%%%%%%%%
\newcommand{\be}{\begin{equation}}\newcommand{\ee}{\end{equation}}%
\newcommand{\bd}{\begin{displaymath}}\newcommand{\ed}{\end{displaymath}}%%%%%%%%%%%
\newcommand{\bit}{\begin{itemize}}                                        %%%%%%%%%
 \newcommand{\eit}{\end{itemize}}                                         %%%%%%%%%
\newcommand{\ben}{\begin{enumerate}}                                      %%%%%%%%%
 \newcommand{\een}{\end{enumerate}}                                       %%%%%%%%%
\newcommand{\baa}{\begin{array}{lll}}                                     %%%%%%%%%
 \newcommand{\eaa}{\end{array}}                                           %%%%%%%%%
\newcommand{\ba}{\begin{eqnarray}}                                        %%%%%%%%%
 \newcommand{\ea}{\end{eqnarray}}                                         %%%%%%%%%
\newcommand{\Ds}{\displaystyle}                                           %%%%%%%%%
\newcommand{\va}[1]{\langle{#1}\rangle}                                   %%%%%%%%%
\newcommand{\gev}[1]{\relax\ifmmode{\text{GeV}^{#1}}                      %%%%%%%%%
                     \else{GeV$^{#1}${ }}\fi}                             %%%%%%%%%
\def\MSbar{\relax\ifmmode\overline                                        %%%%%%%%%
            {\rm MS}\else{$\overline{\rm MS}${ }}\fi}                     %%%%%%%%%
\def\as{\relax\ifmmode \alpha_s\else{$ \alpha_s${ }}\fi}                  %%%%%%%%%
\def\abar{\relax\ifmmode{\bar{a}}\else{$\bar{a}${ }}\fi}                  %%%%%%%%%
\usepackage[greek,english]{babel}
\usepackage[utf8]{inputenc}

\begin{document}
\thispagestyle{empty}
%\date{\today}
\preprint{\hbox{RUB-TPII-01/2013}\\ }
%\vspace*{-10mm}

\title{Can we understand an auxetic pion-photon transition form factor
       within QCD? BaBar faces Belle\footnote{This work is dedicated to the memory of
       Klaus Goeke---friend and colleague.}}
 \author{N.~G.~Stefanis}
  \email{stefanis@tp2.ruhr-uni-bochum.de}
   \affiliation{Institut f\"{u}r Theoretische Physik II,
                Ruhr-Universit\"{a}t Bochum,
                D-44780 Bochum, Germany\\}

 \author{A.~P.~Bakulev\footnote{deceased}}
   \affiliation{Bogoliubov Laboratory of Theoretical Physics, JINR,
                141980 Dubna, Russia\\}

 \author{S.~V.~Mikhailov}
  \email{mikhs@theor.jinr.ru}
   \affiliation{Bogoliubov Laboratory of Theoretical Physics, JINR,
                141980 Dubna, Russia\\}

\author{A.~V.~Pimikov}
  \email{alexandr.pimikov@uv.es}
   \affiliation{Bogoliubov Laboratory of Theoretical Physics, JINR,
                141980 Dubna, Russia\\}
   \affiliation{Departamento de F\'{\i}sica Te\'orica -IFIC,
                Universidad de Valencia-CSIC, E-46100 Burjassot
                (Valencia), Spain\\}

\date{\today}

\begin{abstract}
A state-of-the-art analysis of the pion-photon transition form factor
is presented based on an improved theoretical calculation that includes
the effect of a finite virtuality of the quasi-real photon in the
method of light-cone sum rules.
We carry out a detailed statistical analysis of the existing
experimental data
using this method and by employing pion distribution amplitudes with
up to
three Gegenbauer coefficients $a_2, a_4, a_6$.
Allowing for an error range in the coefficient $a_6\approx 0$, the
theoretical predictions for $\gamma^*\gamma\to\pi^0$
obtained with nonlocal QCD sum rules are found to be in good agreement
with all data that support a scaling behavior of the transition form
factor at higher $Q^2$, like those of the Belle Collaboration.
The data on $\gamma^*\gamma\to\eta/\eta'$ from CLEO and \babar are also
reproduced, while there is a strong conflict with the auxetic trend of
the \babar data above 10~GeV$^2$.
The broader implications of these findings are discussed.
\end{abstract}
%PACS99 used: Renormalization group evolution of parameters=11.10.Hi
%             Perturbative calculations=12.38.Bx
%             Other nonperturbative calculations in QCD=12.38.Lg
%             Electromagnetic Form Factors=13.40.Gp
%\keywords{Transition form factors,
%          pion distribution amplitude,
%          higher twist,
%          light-cone sum rules,
%          collinear factorization,
%          higher-order radiative corrections,
%          renormalization group evolution}

\maketitle

\section{Introduction}
\label{sec:intro}
The data of the \babar Collaboration \cite{BaBar09} of the
$\gamma^*(q_{1}^{2})\gamma(q_{2}^{2})\to \pi^0$
(with $q_{1}^{2}=-Q^2$ for the far off the mass shell photon and
$q_{2}^{2}=-q^2\approx 0$ for the near on mass shell photon)
transition form factor (TFF) in the wide momentum-transfer range
from 4 to 40 GeV$^2$ have not yet found a satisfactory explanation
within the (collinear) factorization approach of QCD.
As first pointed out in \cite{MS09}, the rise of the scaled form factor
$
 Q^2F^{\gamma^{*}\gamma\pi^0}(Q^2)
\equiv
 \mathcal{F}^{\gamma\pi}(Q^2)
$
observed by \babar above 10~GeV$^2$ up to the highest momentum probed
(with the exception of two data points at about 14 and 27~GeV$^2$ that
are below and close to the asymptotic limit
$\mathcal{F}^{\gamma\pi}_{Q^2\to\infty}(Q^2)=\sqrt{2}f_\pi$~GeV with
$f_\pi \approx 0.131$~GeV)
does not conform with the standard QCD approach based on collinear
factorization---see \cite{BL89} for a review.

Following the publication of the \babar data on
$\mathcal{F}^{\gamma\pi}(Q^2)$
in 2009 there was a spurt of worldwide theoretical activity, using
different approaches and drawing strongly diverging conclusions
(see \cite{BMPS11} for the original analysis and \cite{SBMP11} for a
brief benchmark comparison with other approaches).
For instance, Agaev et al. \cite{ABOP10} claimed that the
\babar data for the pion-photon transition form factor (TFF) can be
accommodated within QCD using light-cone sum rules (LCSR)s in
conjunction with pion distribution amplitudes (DA)s characterized by
an inverse hierarchy of Gegenbauer coefficients
$a_4>a_2$ (and eventually including still higher terms
$a_{6},a_{8}, a_{10},...$).
In contrast, our recent analysis in \cite{BMPS11}---which utilizes
basically the same method but in connection with endpoint-suppressed
pion DAs \cite{BMS01} based solely on $a_2, a_4$---comes
to the opposite conclusion.

In 2012, the Belle Collaboration \cite{Belle12} reported upon
a new measurement of the process $\gamma^*\gamma \to \pi^0$ at the KEKB
collider for the kinematical region
4~GeV$^2 \lesssim Q^2 \lesssim 40$~GeV$^2$.
The main message from this new experiment is that the measured values
of $\mathcal{F}^{\gamma\pi}(Q^2)$
agree with the previous measurements of CELLO \cite{CELLO91} and CLEO
\cite{CLEO98} and also with the data of \babar in the momentum
range
$Q^2\lesssim 9$~GeV$^2$,
while at still higher momenta they do not show a growth with $Q^2$ but
are more or less close to the asymptotic limit of QCD with the
exception of a single point that gives a larger value of
$\mathcal{F}^{\gamma\pi}(Q^2)$.
Surprisingly, this outlier at 27~GeV$^2$ shows exactly the opposite
behavior relative to the \babar measurement at the same momentum value
that coincides with the asymptotic limit.

Such an incongruent behavior of the data does not allow a unique
theoretical description, because there is no characteristic mathematical
signature which emerges from the statistics of these measurements that
would allow to draw reliable conclusions about the size of the scaled
TFF at large $Q^2$.
This issue was pondered in our recent paper in \cite{BMPS12}, in which
we proposed a basic classification scheme of the available data
juxtaposed with the cutting-edge theoretical predictions from various
approaches.
Referring to the scaled TFF vs. $Q^2$ (see Fig.\ 2 in \cite{BMPS12}),
this scheme consists of two distinct bands---one exhibiting scaling
at high $Q^2$(Belle data),
the other sloping upward (\babar data), while there is a
third band in between collecting some theoretical results that are
indifferent.
Pooling all data in a single database (see Fig.\ 3 in \cite{BMPS12}),
one faces the problem that the underlying theoretical approaches are
hardly compatible to each other because they correspond to very
different underlying mechanisms.

These issues were further investigated in \cite{PBMS12,BMPS12_LC2012},
where we performed a two- and a three-parameter fit to all data
utilizing, correspondingly, the Gegenbauer coefficients
$a_2, a_4$ and $a_2, a_4, a_6$
within LCSRs.
The main conclusion from these studies is that both fits to the
combined sets of
the
data from CELLO, CLEO, and Belle (termed CCBe), on one
hand, and the set consisting of the CELLO, CLEO, and \babar data
(termed CCBB), on the other, have no overlap.
The CCBe band indicates scaling, as predicted by QCD, while the CCBB
band exhibits an auxetic\footnote{This word derives from
the Greek word \foreignlanguage{greek}{a'uxhsis}
which means the inherent tendency to increase.
Auxetic materials have a negative Poisson's ratio and increase their
cross section %become thicker
perpendicular to the applied force when stretched.
That behavior is considered as an oddity.}
behavior that cannot be accommodated within the standard QCD scheme of
collinear factorization.
This finding reinforces our previous results in \cite{BMPS12}, favoring
the classification pattern of two distinct bands rather than a single one
that encompasses all data.
Moreover \cite{PBMS12,BMPS12_LC2012}, the CCBe data set supports the
theoretical predictions derived with the help of LCSRs in
\cite{BMS03,BMS05lat} using a pion DA (termed BMS) extracted before
from QCD sum rules with nonlocal condensates in \cite{BMS01}---see also
\cite{MR86,BR91}.
Indeed, the BMS $\pi^0$ DA fits the CCBe data
in terms of $a_2$ and $a_4$
with an accuracy of
$\chi^2_{\rm ndf}=\chi^2/{\rm ndf}=22.1/33$,
where ndf=number of degrees of freedom.

Moreover, the same calculation \cite{BMPS11} agrees with the \babar
data \cite{BaBar11-BMS} for the processes $\gamma^*\gamma\to\eta$,
$\gamma^*\gamma\to\eta'$ using the description of the $\eta-\eta'$
mixing in the quark flavor basis \cite{FKS98} to relate the form factor
of the $|n\rangle =(1/\sqrt{2})(|u\bar{u}\rangle + |d\bar{d}\rangle)$
state to that of the pion.
An immediate implication of this agreement is that the DA of the
nonstrange component $|n\rangle$ of the $\eta$, $\eta'$ mesons
should be similar in shape to that of the $\pi^0$.
This would mean in turn that there should be no strong flavor symmetry
breaking in the pseudoscalar meson sector of QCD.
On the other hand, the CCBB data set does not support the BMS-type of
DAs, that have their endpoints suppressed, and demands the inclusion of
at least the next coefficient $a_6$, or even higher ones, as
proposed in \cite{ABOP10}.
Best agreement with the CCBB set is, however, provided with flat-type
DAs \cite{Rad09,Pol09} which, at the same time, fail to reproduce the
CCBe data.

In this work, we extend and refine our data analysis for
$\mathcal{F}^{\gamma\pi}(Q^2)$
\cite{BMS02,BMS03,BMS05lat,MS09,BMPS11,BMPS_QCD2011,BMPS12}
(consult also \cite{ABOP10,ABOP12}) in the following points :
(i) We estimate quantitatively the theoretical uncertainty owing to
the small virtuality of the quasi-real photon in an attempt to take
into account the unknown dependence on the momentum transfer to the
untagged electron.\footnote{%
We thank Wojciech Broniowski for attracting our attention to this
point (see in this context also \cite{BA09,ArBr10}).
The dependence of the $\pi^0$ TFF on the virtuality of the quasi-real
photon up to
%$|q_{2}^{2}|=0.2$~GeV$^2$
was recently discussed in
\cite{Lic10}, using the vector-meson-dominance hypothesis.}
To this end, we define a new quantity $\Delta(Q^2)$ that ``measures''
the susceptibility of the TFF to the variations of $q^2$.
(ii)
We extend the original BMS %scheme
DA ``bunch'' by considering the correlated
``noise'' related to the coefficient $a_6$ \cite{BMS01} and allowing
it to vary with an appropriate error rate.
(iii)
We consider in detail the statistical properties of the \babar
and the Belle data and discuss the features of the statistical
fluctuations and their influence on the $Q^2$-behavior of the
pion-photon TFF
in terms of two different fit models used in the literature.
(iv) We give a qualitative discussion of the spacelike TFFs
$
 Q^2F^{\gamma^*\gamma\eta(\eta')}(Q^2)
\equiv
 \mathcal{F}^{\gamma\eta(\eta')}(Q^2)
$
in comparison with the recent data of \babar \cite{BaBar11-BMS}
and the older ones of CLEO \cite{CLEO98}.

The paper is organized as follows: In Sec.\ \ref{sec:theory} we sketch
the theoretical scheme used in this work.
In Sec.\ \ref{sec:analysis} we present a theoretical tool to probe the
sensitivity of the TFF to the small photon virtuality.
The inclusion of the correlated noise owing to the coefficient
$a_6$ is discussed in the same section.
Section \ref{sec:statistics} is devoted to the statistical analysis of
the various data sets relative to each other and against theoretical
predictions.
Section \ref{sec:discussion} contains an in-depth discussion of our
results, while our conclusions are drawn in Sec.\ \ref{sec:concl}.
Important technical details are collected in three appendices.

%%%%%%%%%%%%%%%%%%%%%%%%%%%%%%%%%%%%%%%%%%%%%%%%%%%%%%%%%%%%%%%%%%%%%%% Figure 1
\begin{figure*}[t!]
\centerline{\includegraphics[width=0.40\textwidth]{% Fey_ee2eepi_5.eps}
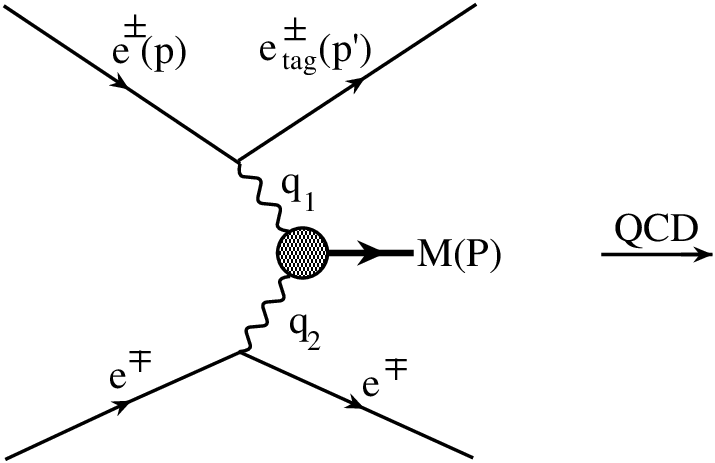}
 \vspace*{2mm}
 ~~~~~~\includegraphics[width=0.40\textwidth]{% Fey_LO_NLO_NNLO_new.eps}}
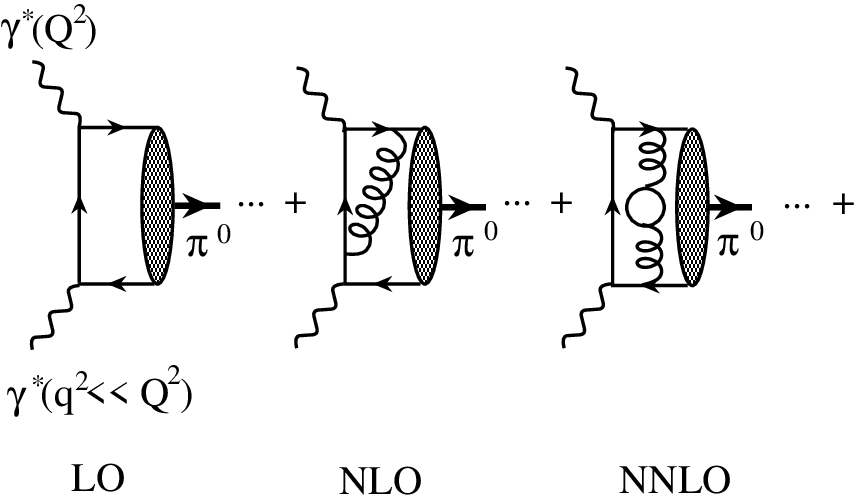}}
\vspace{1mm}
\caption{Left. Generic experimental setup for the process
$e^{+}e^{-}\to e^{+}e^{-} M$, where $M$ is a pseudoscalar meson
$\pi^{0}, \eta$ or $\eta'$.
The tagged electron (or positron) is labeled and the corresponding
momenta of all particles are denoted.
Right. Basic diagrams describing the process within QCD on the basis
of collinear factorization order by order of perturbation theory:
leading order (LO), next-to-leading order (NLO),
and next-to-next-to-leading order (NNLO).
\label{fig:exp-setup}
\vspace{-3mm}}
\end{figure*}
%%%%%%%%%%%%%%%%%%%%%%%%%%%%%%%%%%%%%%%%%%%%%%%%%%%%%%%%%%%%%%%%%%%%%%%

\section{Theoretical basis}
\label{sec:theory}

The pion-photon TFF $F^{\gamma^*\gamma^*\pi^0}$
is described by the following matrix element
\begin{eqnarray}
&& \int\! d^{4}z\,e^{-iq_{1}\cdot z}
  \langle
         \pi^0 (P)| T\{j_\mu(z) j_\nu(0)\}| 0
  \rangle
=
  i\epsilon_{\mu\nu\alpha\beta}
  q_{1}^{\alpha} q_{2}^{\beta}
\nonumber \\
&&  ~~~~~~~~~~~~~~ \times ~
  F^{\gamma^{*}\gamma^{*}\pi^0}(Q^2,q^2)\ ,
\label{eq:matrix-element}
\end{eqnarray}
%Eq (1)
where $j_\mu$ is the quark electromagnetic current.
For a far-off-shell photon with the momentum $Q^2$ and an almost real
photon with the small virtuality $q^2$ the TFF within the method of
LCSRs is given by the expression
\begin{widetext}
\begin{eqnarray}
  Q^2 F^{\gamma^*\gamma^*\pi}\left(Q^2,q^2\right)
=
  \frac{\sqrt{2}}{3}f_\pi
  \left[
        \frac{Q^2}{m_{\rho}^2+q^2}
        \int_{x_{0}}^{1}
        \exp\left(
                  \frac{m_{\rho}^2-Q^2\bar{x}/x}{M^2}
            \right)
        \bar{\rho}(Q^2,x)
  \frac{dx}{x}
  +    \int_{0}^{x_0} \bar{\rho}(Q^2,x)
        \frac{Q^2dx}{\bar{x}Q^2+x q^2}
  \right]
\, .
\label{eq:LCSR-FQq}
\end{eqnarray}
\end{widetext}
%Eq (2)
Here $x$ denotes the longitudinal momentum fraction carried by the
struck parton,
$\bar{x}=1-x$
is the momentum of the spectator,
$\Ds x_0=(Q^2)/\left(Q^2+s_0\right)$, $s =\bar{x}Q^2/x$,
$M^2$ is the Borel parameter, and the spectral density has the
form
$
 \bar{\rho}(Q^2,x)=(Q^2+s)\rho^\text{pert}(Q^2,s)
$,
where
\begin{eqnarray}
  \rho^\text{pert}(Q^2,s)
& = &
  \frac{1}{\pi} {\rm Im}F^{\gamma^*\gamma^*\pi^0}
  \left(Q^2,-s-i\varepsilon\right)
\nonumber \\
& = &
   \rho_\text{tw-2}
  +\rho_\text{tw-4}
  +\rho_\text{tw-6}
  +\ldots\, .
\label{eq:rho-twists}
\end{eqnarray}
%Eq (3)
Each term of definite twist (abbreviated by tw) is calculable
with the help of the analogous hard part convoluted with the pion
distribution amplitude of the same twist \cite{Kho99}.

The Taylor expansion of the TFF with respect to
$q^2$ to order $\delta q^2$ reads
\begin{widetext}
\begin{eqnarray}
  Q^2 F^{\gamma^*\gamma\pi}\left(Q^2,q^2\right)\Big|_{q^2=0}
  + Q^2 \frac{d}{dq^2} F^{\gamma^*\gamma \pi^0}(Q^2, q^2)\Big|_{q^2=0}\delta q^2
\equiv
  \mathcal{F}^{\gamma\pi}(Q^2)
  + \left(\mathcal{F}^{\gamma\pi}\right)'_{q^2=0}(Q^2)\delta q^2\, .
  \label{eq:LCSR-Teylor}
\end{eqnarray}
\end{widetext}
%Eq (4)
The first term in (\ref{eq:LCSR-Teylor}) is the usual expression
obtained with a strictly real photon, whereas the second one gives the
first non-vanishing contribution of a finite virtuality.
The sum of the twist-two and the twist-four spectral densities,
$\bar{\rho}(Q^2,x)$,
as well as the explicit form of the terms on the r.h.s. of
Eq.\ (\ref{eq:LCSR-Teylor}) are provided in Appendices
\ref{sec:FFTs} and \ref{sec:H-amplitude}, respectively.
In this work, the twist-four term is taken into account in the
effective form \cite{Kho99}
$
 \varphi_{\pi}^{(4)}(x,\mu^2)
\sim
 \delta_\text{tw-4}^2(\mu^2)\,x^2(1-x)^2
$.
As usual in LCSR calculations, $s_0\simeq 1.5~\text{GeV}^2$ is the
effective threshold in the vector channel.

At the level of twist-two, the TFF has the following expansion
\begin{eqnarray}
  F_{\gamma^*\gamma^*\pi^0}^\text{tw-2}
& \sim &
  \left[
  T_\text{LO}
 +a_s(\mu^2) T_\text{NLO}
 +a_s^2(\mu^2) T_{\text{NNLO}_{\beta_0}}+\ldots
 \right]
\nonumber \\
& & \! \otimes ~
  \varphi_{\pi}^\text{(2)}(x,\mu^2)\, ,
\label{eq:T_exp}
\end{eqnarray}
%Eq (5)
in which the leading- (LO), next-to-leading (NLO), and
next-to-next-to-leading (NNLO) terms are displayed.
The corresponding Feynman graphs are depicted in
Fig.~\ref{fig:exp-setup}.
Note that our calculations here are incorporating the NLO spectral
density in the corrected form pointed out in \cite{ABOP10}.
The nonperturbative content of the TFF is encoded in the pion DA.
In our previous analysis in \cite{BMPS11,BMPS12}, we have considered
several proposed models for $\varphi_{\pi}^{(2)}$ and have compared
the predictions extracted from them with all existing experimental
data.
For the scope of the present analysis, it is sufficient to employ
only the ``bunch'' of twist-two pion DAs determined in \cite{BMS01}
within the framework of QCD sum rules (SR) with nonlocal condensates
(NLC).
This DA ``bunch'' can be effectively parameterized in terms of the
first two Gegenbauer coefficients $a_2$ and $a_4$, so that one has
($\xi \equiv x-\bar{x}$)
\begin{equation}
  \varphi_{\pi}^\text{BMS}(x)
=
6x\bar{x}
 \left[
       1 + a_2 C_{2}^{3/2}(\xi) + a_4 C_{4}^{3/2}(\xi)
 \right],
\end{equation}
%Eq (6)
whereas all higher coefficients have also been determined but found
to be negligible, albeit with relatively large uncertainties,
so that they were ignored.
The DA with the coefficients
$a_2(\mu^2)=0.20$
and
$a_4(\mu^2)=-0.14$ (at $\mu^2=1$~GeV$^2$)
is termed the BMS model (from Bakulev, Mikhailov, Stefanis)
and is marked out by the fact that its first ten moments
$\langle \xi^{N} \rangle_{\pi}
\equiv
  \int_{0}^{1} dx (2x-1)^{N} \varphi_{\pi}^{(2)}(x,\mu^2)$
with the normalization condition
$\int_{0}^{1} dx \varphi_{\pi}^{(2)}(x, \mu^2)=1$
lie within a particular range of values computed in
\cite{BMS01}.
The connection between the moments
$\langle \xi^{N} \rangle_{\pi}$
and the coefficients $a_{2n}$ is outlined in
Appendix \ref{sec:mom-coef}.
Covering
\textit{the whole} admissible set of $\{a_2,~a_4\}$
values gives rise to the ``BMS bunch''
of pion DAs.
The extension of this ``bunch'' to a 3D set of
coefficients $\{a_2,~a_4,~a_6\}$ will be considered further below.

The key characteristic of the BMS $\pi^0$ DAs is that their
kinematic endpoints $x=0,1$ are strongly suppressed.
This suppression is related to the assumption that the vacuum
quarks have a non-zero virtuality
$
 \lambda_{q}^2
=
 \qq{\left(igG_{\mu\nu}\sigma_{\mu\nu}\right)}/2\qq{}
\simeq
 (0.35-0.55)$~GeV$^2$, pertaining to the use of QCD
SRs with NLCs \cite{MR86,BR91,BMS01}.
This is an approach rooted in the hypothesis that in the coordinate
representation the NLCs are not constant but depend on the Euclidean
separation of the quark fields and decay with a correlation length
$\Lambda_{\rm corr} \sim 1/\lambda_q$---see \cite{BMS04kg,BM02} for
more details and \cite{SSK99} for related explanations.
Then, the endpoint contributions in the scalar condensate that
dominates the pion sum rule are strongly suppressed---in contrast to
the standard approach of Chernyak and Zhitnitsky (CZ) \cite{CZ84},
in which just these regions dominate the $\pi^0$ DA.

It turns out \cite{BMPS11,BMPS12} that the majority of the existing
experimental data at intermediate $Q^2\leq 9$~GeV$^2$ are best
described by such endpoint-suppressed pion DAs \cite{BMS01}.
This becomes obvious from Fig.\ \ref{fig:pionFF-strips}
(upper green band), where we compare our predictions with
various experimental data using for the horizontal axis a logarithmic
scale in order to ``stretch out'' the smaller values of $Q^2$.
In particular the Belle data are in good agreement with our
predictions up to 40~GeV$^2$, whereas the high-$Q^2$ tail of the
$\pi^0$ \babar data is in conflict with these predictions.

\section{Methods of data analysis}
\label{sec:analysis}

\subsection{Small virtuality of quasi-real photon}
\label{subsec:fin-virt}

In the first part of this section we address the treatment of a small
but finite virtuality of the quasi-real photon within the method of
LCSRs.

\begin{figure}[t!]
 \centerline{\hspace{0mm}\includegraphics[width=0.48\textwidth]{% fig-BMSvsData.q2ErM2.004.eps}}
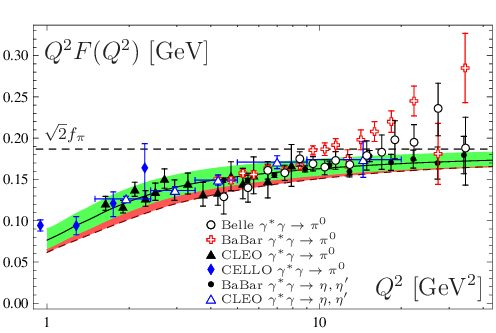}}
  \caption{\label{fig:pionFF-strips} (color online).
    Logarithmic plot of the theoretical predictions for the scaled
    $\gamma^*\gamma\pi^0$
    transition form factor in comparison with data taken from various
    experiments, as indicated.
    The upper (green) band shows the results obtained
    within our approach \protect\cite{BMPS11,BMPS12} assuming that the
    quasi-real photon has vanishing virtuality.
    The lower (red) strip represents the influence
    on the TFF of the small
    virtuality of the quasi-real photon induced by the untagged
    electron in the Belle experiment ($q^2 \approx 0.04$~GeV$^2$).
    }
\end{figure}

As we announced in the Introduction and expounded in Sec.\
\ref{sec:theory}, we include into our calculation of the $\pi$ TFF
the small virtuality of the quasi-real photon in an attempt to mimic
the real situation of a single-tag experiment, like that of \babar
and Belle.
Such experiments  bear an uncertainty owing to the unknown dependence
on the momentum transfer to the untagged electron.
This means that the facility can register only events with a
momentum of the quasi-real photon $|q^{2}|$ up to some limiting
value.
To get an estimate of this effect on the calculation, we use for our
numerical evaluation the area of photon momenta probed in
the Belle experiment \cite{Belle12}
and allow a variation of $q^2$ in the range
$q^2 \lesssim 0.04$~GeV$^2$ down to 0.01~GeV$^2$%
\footnote{
The value $q^2 \simeq 0.04$~GeV$^2$ is due to S. Uehara,
private communication during the Meson TFF Workshop, 29-30 May, 2012,
Cracow, Poland.}.
This area is compatible with the cuts imposed by CLEO which accept
two charged tracks, each of transverse momentum of the order of
$\delta q^2\simeq0.01$~GeV$^2$,
while \babar \cite{BaBar09} imposed
an upper limit of $|q_{2}^{2}| < 0.18$~GeV$^2$.
Employing Eqs.\ (\ref{eq:LCSR-FQq}) and (\ref{eq:LCSR-derF}), we compute
the TFF and present the result graphically in
Fig.\ \ref{fig:pionFF-strips} in terms of the lower narrow (red)
strip that enlarges the original BMS green band downwards.
%%%%%%%%%%%%%%%%%%%%%%%%%%%%%%%%%%%%%%%%%%%%%%%%%%%%%%%%%%%%%%%%%%%%%%% Figure 2
%%%%%%%%%%%%%%%%%%%%%%%%%%%%%%%%%%%%%%%%%%%%%%%%%%%%%%%%%%%%%%%%%%%%%%%

The other key ingredients of our LCSR calculation
in this work are the following:
QCD radiative corrections with NLO accuracy and the
twist-four contribution are explicitly included.
The main NNLO term, proportional to $\beta_0$ \cite{MMP02}, is taken
into account implicitly by means of uncertainties together
with the twist-six term calculated in \cite{ABOP10}.
Note that the NLO, NNLO$_{\beta}$, and twist-four contribution are
all negative, supplying suppression, whereas the twist-six
term has a positive sign and is either very small if a Borel parameter
of $M^2=1.5$~Gev$^2$ is used---as in \cite{ABOP10}---or it has for
$M^2$ varying in the interval
$M^2\in[0.7;0.9]$~GeV$^2$---as in \cite{BMPS11,BMPS12}---approximately
the same size as the NNLO$_{\beta}$ radiative correction and
almost cancels against it.
It is worth mentioning that the Borel parameter in our approach
is not fixed to a particular value, but is allowed to vary with $Q^2$
according to the relation
$M^2=M_\text{2-pt}^2/\va{x}(Q^2)$, i.e., becoming smaller with increasing
$Q^2$.
Here $M_\text{2-pt}^2$ is the two-point Borel parameter that is
specified in the two-point QCD SR for the $\rho$-meson at the mean value
$M_\text{2-pt}^2=0.7$~GeV$^2$,
while $\va{x}(Q^2)$ is some average value of $x$ at fixed value of $Q^2$.
We emphasize that our results are not particularly sensitive to this
treatment of $M^2$.
Indeed, the difference of the TFF results obtained with $M^2(Q^2)$ relative
to those computed with a fixed $M^2$ value varies in the $Q^2$ range
$[1\div 40]$~GeV$^2$ from $-2\%$ for $M^2=0.7$~GeV$^2$ to $+3\%$ to
the maximum of $M^2=0.9$~GeV$^2$.
Would we set $M^2=1.1$~GeV$^2$, a value well outside the interval
mentioned above, the influence on the TFF in the relevant region of
$Q^2$ between 10~GeV$^2$ and 40~GeV$^2$, in which our TFF predictions
for the BMS ``bunch'' almost scale, would be not more than $4\%$.

The leading-twist pion DA entering the LCSRs is determined in the
framework of QCD SR NLCs \cite{BMS01}.
Additional suppression results from the evolution of the Gegenbauer
coefficients in the parameterization of the pion DA, taken into
account in our analysis at the NLO level.
The theoretical uncertainty entailed by the small photon
virtuality accumulates as suppression expressed in the form of
the narrow (red)
strip below our original TFF predictions
\cite{BMPS11} in Fig.\ \ref{fig:pionFF-strips}.
This additional uncertainty somewhat increases
at lower $Q^2$ values
the width of the upper (green) band.
The latter contains the
variation of the shape of the pion DA
in terms of $a_2, a_4$
within the framework of QCD SR NLCs \cite{BMS01} in conjunction with
uncertainties owing to the twist-four coupling
$\delta_\text{tw-4}^2(\mu^2)=0.19\pm 0.04$ GeV$^2$ \cite{BMS02}.

%%%%%%%%%%%%%%%%%%%%%%%%%%%%%%%%%%%%%%%%%%%%%%%%%%%%%%%%%%%%%%%%%%%%%%% Figure 3
\begin{figure}[th!]
 \centerline{\hspace{0mm}\includegraphics[width=0.48\textwidth]{% fig-DerLogFF.Asy.CZ.BMS.Belle.eps}}
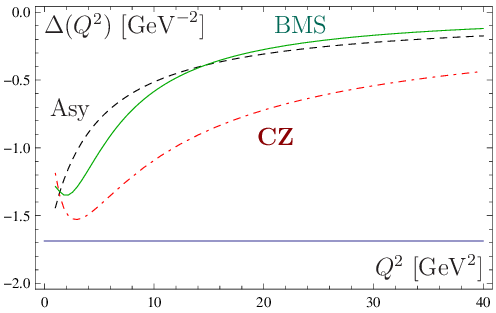}}
  \caption{\label{fig:delta.Q2} (color online).
    Parametrization of the dependence of the TFF on the small but finite
    photon virtuality defined in Eq.\ (\ref{eq:delta.Q2}) in terms of
    $\Delta(Q^2)$ for three characteristic pion DAs:
    CZ \protect\cite{CZ84} --- dashed-dotted (red) line,
    BMS \cite{BMS01} --- solid (green) line,
    Asy --- dashed line.
    The solid horizontal line corresponds to the
    model in \cite{Belle12} that provides $\Delta=1/m_\rho^2$.
    }
\end{figure}
%%%%%%%%%%%%%%%%%%%%%%%%%%%%%%%%%%%%%%%%%%%%%%%%%%%%%%%%%%%%%%%%%%%%%%%
To confront theory with single-tag experiments more precisely, we define the
susceptibility (linear response)
\begin{equation}
  \Delta(Q^2)
\equiv
\frac{\left(\mathcal{F}^{\gamma\pi}\right)'_{q^2=0}(Q^2)}{\mathcal{F}^{\gamma\pi}(Q^2)}
\label{eq:delta.Q2}
\end{equation}
%Eq (7)
which describes the relative sensitivity of the TFF to the variation of
$q^2$.
Thus, the relative change induced by a small virtuality of the
quasi-real photon is
$  \delta\mathcal{F}^{\gamma\pi}/\mathcal{F}^{\gamma\pi}
=
  \Delta(Q^2)\cdot \delta q^2
$,
where the first factor represents the theoretical prediction,
which contains the effects of strong interactions,
and the second one is set by experiment.
As a result, the TFF that includes the small-virtuality effect reads
\begin{equation}
  \tilde{\mathcal{F}}^{\gamma\pi}(Q^2,\delta q^2)
=
  \mathcal{F}^{\gamma\pi}(Q^2)
         \left[
               1 + \Delta(Q^2) \delta q^2
         \right] \ .
\end{equation}
%Eq (18)
Note that $\Delta(Q^2)$ depends on the shape of the pion DA employed
in the calculation.
This dependence is shown in Fig.\ \ref{fig:delta.Q2} for the asymptotic
(Asy) DA,
and the
BMS \cite{BMS01}, and Chernyak-Zhitnitsky (CZ) \cite{CZ84} models.
One observes from this figure that $\Delta(Q^2)$ has always a negative
sign and provides suppression to
$\tilde{\mathcal{F}}^{\gamma\pi}(Q^2,\delta q^2)$,
starting at a maximum value of
-5.4\% (-1.3\%) at $Q^2=2$~GeV$^2$
and rapidly decreasing to
-0.5\% (-0.1\%) for $Q^2 > 30$~GeV$^2$, where we have used the
values
$\delta q^2=0.04 (0.01)$~GeV$^2$ related to the Belle experiment
\cite{Belle12}.
The horizontal (blue) line in Fig.\ \ref{fig:delta.Q2} shows
an artificially reinforced constant response $\Delta$ obtained
with a particular model used in \cite{Belle12}.
Note that similar observations were done in Ref.\ \cite{CIKS12}
using Monte Carlo simulations in which a kinematic cut of
$|q_{2}^{2}| < 0.18$~GeV$^2$ related to \babar ~\cite{BaBar09}
was imposed and
found to lead to a reduction of the cross section at the level of
$3\%$ for the whole $Q^2$ range from 1 to 35~GeV$^2$.

The main conclusions from these findings are twofold:
First, a more precise comparison of theoretical predictions with the
experimental data should account for the final virtuality of the
quasi-real photon because it induces a non-negligible effect.
Second, any calculation with a QCD-based $\pi$ DA model will receive
additional suppression so that the chances to reconcile theoretical
predictions with the \babar data will decrease even further,
with the asymptotic DA loosing ground against all existing data.
In contrast, the enlarged band of our theoretical predictions, obtained
from the BMS ``bunch'' of $\pi$ DAs (Fig.\ \ref{fig:pionFF-strips}) still
includes all CLEO data with their error bars and most of the Belle data,
while even the \babar data below 10~GeV$^2$ are also covered.

\subsection{3D pion DA models from NLC QCD SRs}
\label{sec:3DDA}

%%%%%%%%%%%%%%%%%%%%%%%%%%%%%%%%%%%%%%%%%%%%%%%%%%%%%%%%%%%%%%%%%%%%%%% Figure 4
\begin{figure}[b!]
 \centerline{\hspace{0mm}\includegraphics[width=0.45\textwidth]{% fig-CCBa.CCBe.BMS3D.3D.eps}}
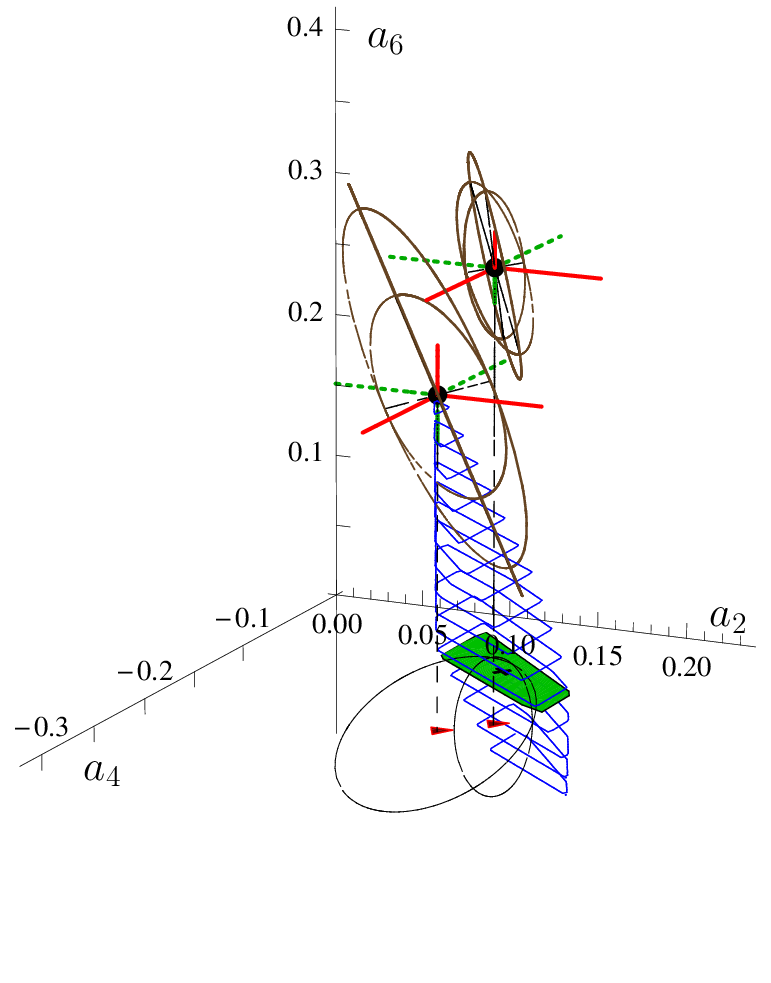}}
  \vspace*{-3mm}\caption{\label{fig:BMS.3D} (color online).
  3D graphics of the pion DA ``bunch''
  obtained from QCD SRs with NLCs, in terms of the coefficients
  $a_2, a_4, a_6$,
  shown as a flight of ``stairs'' of slanted rectangles,
  while the original BMS ``bunch'' in the plane $(a_2,a_4)$
  is shown as a (green) rectangle.
  The displayed $1\sigma$-error ellipsoids represent fits to
  two data sets: smaller ellipsoid (CCBB) \protect\cite{CELLO91,CLEO98,BaBar09}
  and larger ellipsoid (CCBe) \protect\cite{CELLO91,CLEO98,Belle12}.
  The theoretical $\Delta \delta^2_\text{tw4}$ errors for an
  increasing value of $\delta^2_\text{tw4}$ are indicated by a solid
  (red) hairline cross in the forefront, whereas a dashed (green)
  hairline cross in the background denotes a decreasing value.
  All displayed results were calculated at the scale $\mu=2.4$~GeV,
  with more explanations being given in Sec.\ \ref{sec:3DDA}.
  }
\end{figure}
%%%%%%%%%%%%%%%%%%%%%%%%%%%%%%%%%%%%%%%%%%%%%%%%%%%%%%%%%%%%%%%%%%%%%%%

As mentioned above, the ``BMS'' bunch of pion DAs is based on the first
two (nontrivial) Gegenbauer harmonics.
The corresponding coefficients $a_2$ and $a_4$ were derived from the
first five
$\langle \xi^{2n}\rangle$
moments, estimated in \cite{BMS01,BM98} using
QCD SRs with NLCs.
It was found that the next coefficients $a_6, a_8, a_{10}$, obtained
this way, can be set equal to zero, though they bear rather large
uncertainties.
On the other hand, the
pion DAs constructed with the minimal subset
$(a_2,a_4)$ provide a sufficiently good description of
different pion observables---see \cite{BMS04kg} for a review.
Motivated by the high-$Q^2$ Belle data that are not adequately
described with only two Gegenbauer coefficients \cite{BMPS12_LC2012},
we include into our analysis of the data the next higher term
$a_6 \approx 0.038 \sim 0$
with its associated uncertainties ranging within $[-0.186\div 0.263]$
in the sense of correlated ``noise'', but disregard still higher terms.
The outcome of this procedure is displayed in Fig.\ \ref{fig:BMS.3D}
in which the inclusion of $a_6$ with increasing values is illustrated
as a 3D flight of ``stairs'' of slanted rectangles.

%%%%%%%%%%%%%%%%%%%%%%%%%%%%%%%%%%%%%%%%%%%%%%%%%%%%%%%%%%%%%%%%%%%%%%% Figure 5
\begin{figure}[h!]
 \centerline{\hspace{0mm}\includegraphics[width=0.45\textwidth]{% fig-BMS.3DvsData.Log.eps}}
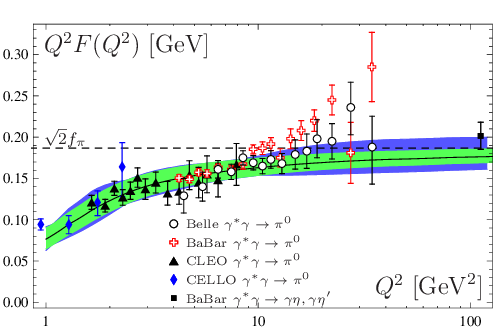}}
  \vspace*{-3mm}
   \caption{\label{fig:BMS.3D.FF} (color online).
    The broader (blue) band, enclosing the narrower (green) one, displays
    the theoretical predictions for the scaled
    $\gamma^*\gamma\pi^0$ TFF obtained with the LCSR approach and pion DAs
    extracted from the $\langle \xi^n \rangle$-moments from \protect\cite{BMS01}
    using three Gegenbauer coefficients $a_2, a_4, a_6$.
    The narrower (green) strip reproduces the results of the original ``BMS
    bunch'', which is shown in Fig.\ \ref{fig:BMS.3D}
    as a shaded slanted rectangle in terms of $a_2$ and $a_4$.
    For comparison, experimental data from various collaborations
    with the indicated labels are also shown.
    }
\end{figure}
%%%%%%%%%%%%%%%%%%%%%%%%%%%%%%%%%%%%%%%%%%%%%%%%%%%%%%%%%%%%%%%%%%%%%%%
The axis of this ``stairs'' incidentally crosses the center of the
CCBe ellipsoid (larger ellipsoid in the forefront).
One observes that
there is no way to satisfy the theoretical constraints
(``stairs'' of slanted rectangles) and the CCBB set of the data
(smaller ellipsoid in the background).

The predictions for the TFF obtained with the three-parametric
pion DA ``bunch'' are shown in Fig.\ \ref{fig:BMS.3D.FF}
by means of a broader (blue) band enveloping the original
(green) one.
One appreciates that the width of the (blue) enveloping band
becomes larger above $Q^2\approx 10$~GeV$^2$, while its
deviation from the original (green) strip below that scale is marginal.
The bottom line is that the DAs from our new 3D ``bunch''
achieve a rather good agreement with the CCBe data at the expense of
larger error bars of the calculated TFF, while even this enlarged band
of predictions is conflicting with the \babar data above
$Q^2\simeq 10$~GeV$^2$.

\section{Statistical data analysis}
\label{sec:statistics}
%%%%%%%%%%%%%%%%%%%%%%%%%%%%%%%%%%%%%%%%%%%%%%%%%%%%%%%%%%%%%%%%%%%%%%% TABLE I
\begin{table*}[th!]
\begin{center}
\begin{ruledtabular}
\caption{Results of the statistical analysis of the \babar
(abbreviated by $\mathfrak{B}$) \protect\cite{BaBar09},
the Belle \protect\cite{Belle12} (abbreviated by $\mathfrak{b}$),
and CELLO\&CLEO data \protect\cite{CELLO91,CLEO98} (abbreviated by
$\mathfrak{CC}$) in terms of a dipole fit (denoted D) and a power-law fit
(denoted P), gauging the accuracy by a goodness of fit $\chi^2_{\rm ndf}$ and
the units of the corresponding $\sigma$, as described in the text.
\label{tab:table-1}}
\smallskip
\smallskip
\begin{tabular}{lcllccc}
Symbol/Name&Fit &\multicolumn{2}{c}{Best fit values}& $\chi^2$ & Relative $\chi^2$ & $\sigma$ deviation \\
\hline
\RedTn{\ding{57}}/$\mathfrak{B}$ & \text{D} & $B_{\mathfrak{B}}=0.23$  & $C_{\mathfrak{B}}$=2.6       & $\chi^2(\mathfrak{B},{\rm D})=1.7$ & $\chi^2(\mathfrak{B} \to \mathfrak{b}, {\rm D})=1.5$  & $\Delta_{\mathfrak{B} \to \mathfrak{b}, {\rm D}}=7.2$
\\
~~$\mathfrak{B}$ & P & $A_{\mathfrak{B}}$=0.182 & $\beta_{\mathfrak{B}}$=0.25  & $\chi^2({\mathfrak{B},{\rm P}})=1.0$ & $\chi^2({\mathfrak{B}\to \mathfrak{b}, {\rm P}})=1.7$ & $\Delta_{\mathfrak{B}\to \mathfrak{b}, {\rm P}}=6.0$
\\
\hline
\ding{109}/{$\mathfrak{b}$}& D & $B_{\mathfrak{b}}$=0.212 & $C_{\mathfrak{b}}$=2.4       & $\chi^2({\mathfrak{b}, {\rm D}})=0.4$ & $\chi^2({\mathfrak{b}\to \mathfrak{B}, {\rm D}})=5.5$ & $\Delta_{\mathfrak{b} \to \mathfrak{B}, {\rm D}}=3.3$
\\
~~$\mathfrak{b}$ & P & $A_{\mathfrak{b}}$=0.169 & $\beta_{\mathfrak{b}}$=0.19 & $\chi^2(\mathfrak{b}, {\rm P})=0.4$ & $\chi^2(\mathfrak{b}\to \mathfrak{B}, {\rm P})=3.7$  & $\Delta_{\mathfrak{b} \to \mathfrak{B}, {\rm P}}=3.7$
\\
\hline
\hline
$\blacktriangle$/$\mathfrak{CC}$ & $\text{D}$ & $B_{\mathfrak{CC}}$=0.176  & $C_{\mathfrak{B}}$=0.82       & $\chi^2({\mathfrak{CC}, {\rm D}})=0.6$ & $\chi^2({\mathfrak{CC}\to \mathfrak{b}, {\rm D}})=1.0$  & $\Delta_{\mathfrak{CC}\to \mathfrak{b},{\rm D}}=8.7$
\\
\BluTn{\ding{108}}/{\rm BL}&D   &$B_{\rm BL}$=0.187  &$C_{\rm BL}$=0.69  & -- &-- &--
\\
\textcolor{darkgreen}{\ding{116}}/\text{LCSR}&D &$B_{\rm LCSR}$=0.180  &$C_{\rm LCSR}$=1.00  & -- &-- &--
\\
\end{tabular}
\end{ruledtabular}
\end{center}
\end{table*}
%%%%%%%%%%%%%%%%%%%%%%%%%%%%%%%%%%%%%%%%%%%%%%%%%%%%%%%%%%%%%%%%%%%%%%%

Before we continue with the statistical analysis of the real data,
it is advisable to state what we should expect from the point of view
of QCD.
If we believe that QCD is the correct microscopic theory of strong
interactions, then it is reasonable to suppose that the data would
cluster with increasing $Q^2$ more and more closely around the
limiting value $\sqrt{2}f_\pi$ becoming equal to that value
(within error bars), if a fictitious experiment were continued
to remote momentum scales.

The antithetic trend between the \babar \cite{BaBar09} and the Belle
\cite{Belle12} data for the $\gamma^*\gamma\pi^0$ TFF calls for a
careful statistical evaluation.
The crucial question is whether these data sets are mutually
supportive or exclusive.
In other words, can we \emph{predict} the trend of the Belle data
using as learning input the \babar data and vice versa?
To answer this question, we use two different parameterizations:
a dipole fit \cite{CLEO98,Belle12}
\begin{equation}
  Q^2|F(Q^2)|
=
  \frac{BQ^2}{Q^2+C}
\label{eq:dipole-fit}
\end{equation}
%Eq (9)
and a power-law fit \cite{BaBar09,Belle12}
\begin{equation}
  Q^2|F(Q^2)|
=
  A
   \left(\frac{Q^2}{10~{\rm GeV}^2}
   \right)^\beta \, ,
\label{eq:powerlaw-fit}
\end{equation}
%Eq (10)
where $B, C, A, \beta$ are free fit parameters.
Both types of fitting functions have been used by
experimentalists \cite{CLEO98,BaBar09,Belle12} before
because of their convenience.
However, we could equally well use another fitting function---it
doesn't really matter.

What matters most is the mutual consistency of such fits in predicting
the trend of the data one from the other.
We use the following convenient abbreviations:
\babar $\equiv \mathfrak{B}$,
Belle $\equiv \mathfrak{b}$,
Dipole $\equiv$ D,
Power-law $\equiv$ P
and express the goodness of fit for each parametrization
in terms of
$\chi^2_\text{ndf}(\text{data set}, \text{fit model})$.
We employ the two mathematical expressions given above to determine the
fit parameters $B,C$ and $A,\beta$ using in turn as input the \babar
($\mathfrak{B}$)
and the Belle ($\mathfrak{b}$) data.
Then, we test how good the obtained fitting model can describe
the other set of data.
The results of this data processing are given in
Table \ref{tab:table-1}.
For future use with respect to fits to the other data sets
CLEO and CELLO,
abbreviated in common by $\mathfrak{CC}$,
we define a relative goodness of fit criterion
$
 \chi^2_{\rm ndf}(\text{data set-1} \to \text{data set-2}, \text{fit model})
$
that serves to explore how the various sets compare to each other., i.e.,
how well the best fit (fit model with best-fit parameters), obtained
from the learning data (set-1), can predict the test data (set-2).
As an example we note
$
 \chi^{2}_{\rm ndf}(\mathfrak{B} \to \mathfrak{b}, D)
$
which describes the fitting of the Belle data from those of
\babar using the dipole formula.
%%%%%%%%%%%%%%%%%%%%%%%%%%%%%%%%%%%%%%%%%%%%%%%%%%%%%%%%%%%%%%%%%%%%%%% Figure 6
\begin{figure*}
\centerline{\includegraphics[width=0.435\textwidth]{% fig-Dipole.Fit.CC_Ba_Be_BMS_BL.eps}
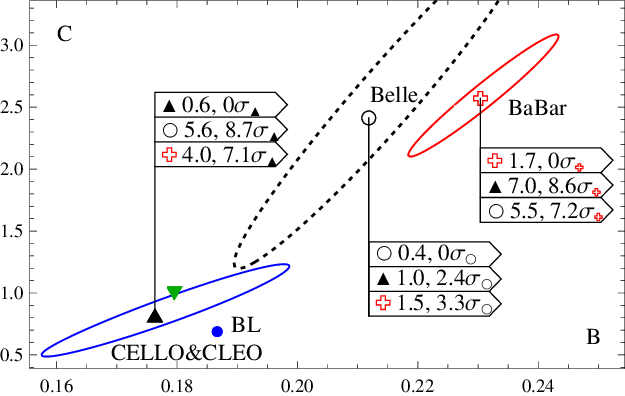}
 \vspace*{2mm}
 ~~~~~~\includegraphics[width=0.45\textwidth]{% fig-Power.Fit.CC_Ba_Be.eps}}
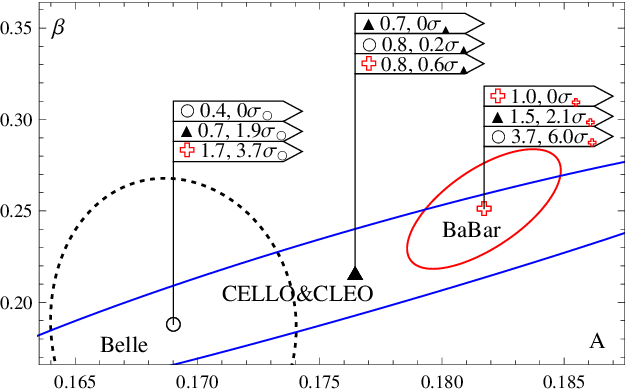}}
\vspace{1mm}
\caption{Left. Dipole fit to the CELLO, CLEO, \babaR and Belle data,
described in terms of the parameters $B$ and $C$, cf.\ Eq.\
(\ref{eq:dipole-fit}).
Right. Analogous graphics for the power-law fit in terms of the
parameters $A$ and $\beta$, cf.\ Eq.\ (\ref{eq:powerlaw-fit}).
More explanations are given in the text.}
\label{fig:fits}
\vspace{-3mm}
\end{figure*}
%%%%%%%%%%%%%%%%%%%%%%%%%%%%%%%%%%%%%%%%%%%%%%%%%%%%%%%%%%%%%%%%%%%%%%%

The strong variation of $\chi^2_{\rm ndf}$ in the upper part
of Table \ref{tab:table-1} reveals that the dynamical behavior with
$Q^2$ of the real Belle data cannot be accurately predicted on the
basis of the \babar data neither with the dipole form nor with the
power-law one.
Indeed the $\chi^2_{\rm ndf}$ values
$\chi^2_{\rm ndf}(\mathfrak{B}\to \mathfrak{b},{\rm D})=1.5$
(dipole fit)
and
$\chi^2_{\rm ndf}(\mathfrak{B}\to \mathfrak{b},{\rm P})=1.8$
(power-law fit)
obtained from the coefficients $(B,C)$ and $(A,\beta)$ via the
\babar data are much larger relative to the value
$
 \chi^2_{\rm ndf}(\mathfrak{b},{\rm D})
=
 \chi^2_{\rm ndf}(\mathfrak{b},{\rm P})=0.4
$,
determined directly from the Belle data.
But also the inverse prediction has not an acceptable precision.
Using the values of $(B,C)$, determined from the Belle data, we find
that the \babar data can be fitted by the dipole fit with a
$\chi^2_{\rm ndf} (\mathfrak{b}\to \mathfrak{B},{\rm D})=5.4$,
while the analogous coefficients $(A,\beta)$ of the power-law fit
would give
$\chi^2_{\rm ndf}(\mathfrak{b}\to \mathfrak{B},{\rm P})=3.7$.

What remains contentious is whether one should prune the outliers
in both data sets.
\babar and Belle made no attempt to explain the origin of the
corresponding outliers, but simply accepted them as a given feature of
their data representing the tails of their probability distribution.
Removing the two \babar outliers, would entail a slightly worse
description of the Belle data. %(lower part of Table \ref{tab:table-1}).
On the other hand, pruning the single outlier of the Belle data would
further improve the scaling behavior of the data with $Q^2$ and
increase the tension to the \babar data.
In both cases, removing the big leaps that are underestimated by the
typical gaussian distribution of each data set, the differences between
successive values of the TFF (called in statistics the ``returns'')
would remain above 10 GeV~$^2$ practically steady and close to zero,
revealing the underlying scaling behavior of the Belle TFF data, while
emphasizing at the same time the auxetic trend of the \babar data in
this $Q^2$ region.

Another important observation from Table \ref{tab:table-1} is that the
%mathematical
particular form of the parametrization used to fit the data is not
crucial.
In fact, the Belle data \cite{Belle12} can be described with both
functional forms---dipole and power-law---with exactly the same
accuracy:
$\chi^2_{\rm ndf} ({\rm b,D})=\chi^2_{\rm ndf}({\rm b,P})=0.4$.
Using another parametrization would not change these findings
significantly.
Thus, without any theoretical presumption, the statistical analysis
of the Belle and the \babar data suggests that they segregate into
two distinct classes of data and cannot merge into a single pool of
aggregated data.
This finding reinforces our conclusions drawn in \cite{BMPS12}
that one should divide the data into two discrete classes
with reference to their $Q^2$ behavior: one showing scaling
(Belle data) and the other exhibiting auxesis (\babar data).

The above discussion can be given a more quantitative meaning by
displaying the precise statistical information linked to each of the
above fits by means of Fig.\ \ref{fig:fits}.
The left panel shows the dipole fit to the CELLO, CLEO, \babaR and
Belle data in terms of the parameters $B$ and $C$, cf.\ Eq.\
(\ref{eq:dipole-fit}), while the right panel contains an analogous
graphics for a power-law fit in terms of the parameters $A$ and $\beta$
(see Eq.\ (\ref{eq:powerlaw-fit})).
We display the $1\sigma$ error ellipses, associated with the
indicated data sets, and mark their centers by a flag which shows the
following values from top to bottom.
First flag: The first number is $\chi^2_\text{ndf}$
for this data set and the second one its
$\sigma$ value.
Second flag:
$\chi^2_{\rm ndf}(\text{marked data set} \to \text{this data set})$
and $\sigma$
with respect to the marked center as seen from the data set
in the first flag;
third flag: in analogy to the previous one but with another
marked center.
The marks for the ellipse centers are displayed in the figure and are
also listed here for convenience:
Cello and CLEO data: $\blacktriangle$;
Belle data: \ding{109};
\babar data: \RedTn{\ding{57}}.
The point labeled by \BluTn{\ding{108}}
corresponds to the Brodsky-Lepage (BL) interpolation formula
(\ref{eq:interpol})
given in the next Section,
while the symbol \textcolor{darkgreen}{\ding{116}}
corresponds to the result of the LCSR calculation which
employs
the BMS DA model.

\section{Discussion of results}
\label{sec:discussion}

In the previous section we concentrated on the statistical analysis of
the experimental data without attempting to provide deeper explanations.
Here we turn to a discussion of the nature and the causes of the
antithetic trends of the present data from the point of view of theory.

There are two rigorous predictions for the behavior of the $\pi^0$ TFF,
one extracted form the axial anomaly in the chiral limit of QCD, i.e.,
at $Q^2=0$ \cite{Adl69,BJ69}, and the other
obtained from QCD in the asymptotic limit $Q^2\to \infty$
\cite{BL81-2phot}.
To interpolate between the $Q^2=0$ and $Q^2=\infty$ limits, the
following phenomenological monopole form has been proposed
Brodsky and Lepage (BL) in
\cite{BL81-2phot}:
\begin{equation}
  F^{\gamma^*\gamma \pi}(Q^2)
=
  \frac{\sqrt{2}f_\pi}{4\pi^2f_\pi^2 + Q^2 } \ .
\label{eq:interpol}
\end{equation}
%Eq (11)
One can derive analogous interpolation formulas for the other
pseudoscalar mesons with $J^{\rm PC}=0^{-+}$, i.e., the $\eta$ and
$\eta'$ in terms of their decay constants $f_\eta$ and $f_{\eta'}$.

Though we lack a detailed theoretical scheme to deal precisely with
nonperturbative QCD, we would expect that the experimental data
would comply with the above QCD preconceptions of the $\pi^0$ TFF.
While this is true for the CLEO and most of the Belle data,
the \babar data indicate a different trend at momenta
$Q^2 \gtrsim 10$~GeV$^2$
that is characterized by a distinctive increase.
Hence, from the QCD point of view, these data appear as being
contingent on unknown enhancement mechanisms of the nonperturbative
quark-gluon interactions.
While we understand the mechanism of endpoint suppression---nonlocal
quark/gluon condensates \cite{BMS01}---we have no clear
understanding of the mechanism of endpoint enhancement which would
give rise to a flat-top pion DA and lead to an auxetic behavior of
the pion-photon transition form factor.
A flat-top pion DA was proposed by Radyushkin~\cite{Rad09}
and in a different context also by Polyakov \cite{Pol09},
while hints for a flat-like pion DA were obtained earlier within
the Nambu--Jona-Lasinio model \cite{AriBro-02} and also in the
so-called Spectral Quark Model \cite{BroArr03}, as well as from the
instanton vacuum \cite{PPRWG99,ADT99}.
Such DAs entail a logarithmic rise of $\mathcal{F}^{\gamma\pi}(Q^2)$
and can indeed comply with the trend of the \babar data for the
$\pi\gamma$ transition.
(see \cite{Dor09,Dor11}).
Meanwhile, several authors have proposed \emph{contextual} explanations
in conjunction with particular low-energy models that can
indeed replicate the growing behavior of
$\mathcal{F}^{\gamma\pi}(Q^2)$
indicated by the \babar data.
Besides the analyses already mentioned, examples are given by the
works in Refs.\
\cite{LiMi09,ArBr10,NV10,WH10,Kro10sud,WH11,PP11,LG12,GZ11,BLM11}.
However, strictly speaking, the flat-top pion DA is an
\textit{after-the-fact} rationalization of the rising scaled TFF
without support from the standard QCD framework.
Thus, it is of little consolation to appeal to contextual
explanations of this effect, though it is possible that some deeper
reason for enhancement may exist---see, for instance,
\cite{Pol09,Dor11,KOT11JETP,ZH11}.

The statistical analysis of the data in this work has shown that the
auxetic trend of the \babar data above 10~GeV$^2$ cannot be predicted
from other experimental data antecedent to them \cite{CELLO91,CLEO98}.
Also the new Belle data~\cite{Belle12} cannot be used to
retrospectively ``predict'' such a behavior of the $\pi\gamma$
TFF using a popular parametrization like the dipole or the
power-law fit (as we have shown in the previous section).
From the theoretical side, the rise of the TFF cannot be intuited
within the standard framework of QCD based on collinear factorization
as well.
Therefore, it is a futile endeavor to try to explain the auxetic
behavior of the \babar data by systematically engineering the fit to
these data as long as there is no deeper understanding of some
underlying dynamical mechanism that should reveal itself
also in other QCD processes.

\begin{figure}[t!]
 \centerline{\hspace{0mm}\includegraphics[width=0.48\textwidth]{% fig-Pred.vs.Data-dot.eps}}
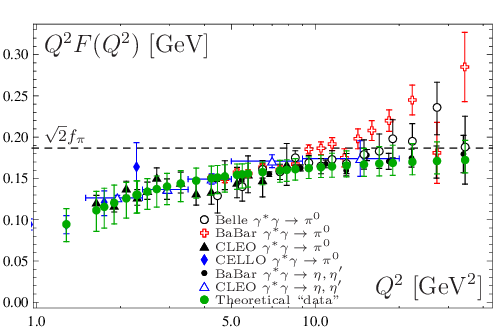}}
  \caption{\label{fig:pionFF-theor-data} (color online).
    Theoretical predictions for the scaled
    $\gamma^*\gamma\pi^0$
    transition form factor in the form of ``theoretical data'',
    including all uncertainties considered in the text, in comparison
    with real experimental data taken from various experiments with
    designations as indicated.
    A logarithmic scale for $Q^2$ is used.
    }
\end{figure}

Comparing our theoretical predictions, computed with the method of
LCSRs and using as nonperturbative input the $\pi$ DAs
extracted from QCD SRs with NLCs, we argue that they provide reasonable
agreement with almost all data---except those of \babar beyond
10~GeV$^2$ (see Fig.\ \ref{fig:pionFF-theor-data}).
Viewed as a function of $Q^2$, the calculated TFF has two
parts in succession: a gentle ascent up to about 10~GeV$^2$, followed
by a saturated part exhibiting scaling beyond that scale in accordance
with perturbative QCD.
Indeed, as one observes from this figure, our predictions shown in
the form of ``theoretical data'' (green bullets with error bars)
comply pretty well with the Belle data within the estimated
uncertainty range, but disagree with the high-$Q^2$ tail of the
\babar data.
At the expense of accepting some ``noise'' for the coefficient
$a_6$, the enlarged error bars of our predictions increase the
agreement with the Belle data significantly, while no reconciliation
with the high $Q^2$ tail of the \babar data is achieved.
This incongruity is divisive in a broader sense because it tells us
that the \babar data are incompatible with scaling of the TFF at large
$Q^2$, a behavior that is a basic characteristic of any
QCD-based calculation.
On the other hand, the high-$Q^2$ trend of the Belle data supports
the scaling behavior of the TFF.
These opposing tendencies cannot be reconciled until more
data will become available in the future.

%%%%%%%%%%%%%%%%%%%%%%%%%%%%%%%%%%%%%%%%%%%%%%%%%%%%%%%%%%%%%%%%%%%%%%% Figure 7
%%%%%%%%%%%%%%%%%%%%%%%%%%%%%%%%%%%%%%%%%%%%%%%%%%%%%%%%%%%%%%%%%%%%%%%

Abstracting from the $\pi^0$ \babar data, an antagonistic mechanism,
governed by quark-gluon strong interactions, that can provide such a
distinctive enhancement to the TFF at large $Q^2$ has yet to be
identified.
One may think that this could eventually be the result of multiple
correlations with various correlation lengths, related to constructive
interference effects, that may prevent partonic interactions governed
by fixed-order or resummed QCD perturbation theory up to excessively
large momentum transfers.
As we have recently argued in \cite{BMPS12}, the antithetic trends of
the \babar and the Belle data, pertaining invariably to auxesis vs.
scaling, correspond to $\pi$ DAs with distinct endpoint
characteristics.
To get a scaling behavior, one needs endpoint suppression but also a
shape that is wider than the asymptotic DA.
As we have shown in this analysis, the asymptotic DA
falls short to comply with all existing data.
On the other hand, endpoint enhancement with only one
but excessively large coefficient $a_2$, like
in the case of the CZ DA, overestimates all data
below 20~GeV$^2$ while being unable at the same time to
reproduce the (at least) logarithmic increase of the
high-$Q^2$ \babar data.
This is only possible if one includes into the $\pi^0$ DA more
coefficients, as proposed in \cite{ABOP10,ABOP12}.
Employing a flat-top $\pi$ DA \cite{Rad09,Pol09}, the agreement with
the \babar data is best but at the expense that one has to
abandon collinear factorization and QCD scaling.

As we discussed in more detail in \cite{BMPS11}, our approach is
capable of capturing the basic features of the
$\gamma^*\gamma\eta(\eta')$ TFF as well.
Using for simplicity the description of the $\eta-\eta'$
mixing in the quark-flavor basis \cite{FKS98} (see \cite{Fel00}
for a review), one has
\begin{equation}
  \begin{pmatrix} |\eta\rangle \\ |\eta'\rangle
  \end{pmatrix}
=
  \begin{pmatrix} \cos\phi & -\sin\phi \\ \sin\phi & \cos\phi
  \end{pmatrix}
  \begin{pmatrix} |n\rangle \\ |S\rangle
  \end{pmatrix} \ ,
\label{eq:mixing}
\end{equation}
%Eq (12)
where the nonstrange part is given by
$|n\rangle=(1/\sqrt{2})(|u\bar{u}\rangle + |d\bar{d}\rangle)$
and the strange component is
$
 |S\rangle
=
 |s\bar{s}\rangle
$,
with the angle $\phi$ denoting the deviation of the mixing angle from
the ideal one owing to the $U_{\rm A}(1)$, i.e., the
axial-vector, anomaly.
Then, the TFFs of the physical $\eta$ and $\eta'$ mesons can be linked
to those of the states $|n\rangle$ and $|S\rangle$---see
\cite{BaBar11-BMS} for further details.
Using the currently accepted value of the mixing angle
$\phi\approx 41^\circ$, as used by the \babar Collaboration
in \cite{BaBar11-BMS}, in order to mix the data on
$\eta$ and $\eta'$, we obtain the data points for the TFF of the
$|n\rangle$ state displayed in Fig.\ \ref{fig:pionFF-theor-data}.
This rough treatment ignores in the $Q^2$ evolution the mixing
with the gluonic components and also the difference in the
normalization owing to the different decay constants, but is
sufficient for our qualitative considerations.

An independent confirmation of the $\gamma^*\gamma\eta(\eta')$
\cite{BaBar11-BMS} data, would establish the agreement with the
asymptotic QCD limit, denoted by the horizontal dashed
line in Fig.\ \ref{fig:pionFF-theor-data}.
It would also agree with our theoretical predictions obtained with the
BMS formalism (``theoretical'' data in the same figure), as we
explained above.
This would mean that (a) the DAs of the two pseudoscalar
mesons $\pi^0$ and $\eta$ (strictly speaking its nonstrange component)
are similar---no (significant) $SU(3)_{\rm F}$ flavor asymmetry---and
(b) have their endpoints strongly suppressed.
Implicitly, this would give support to the idea of NLCs that entail
this suppression and thus validate the sum-rule method in
\cite{BMS01} based on them.
In contrast, the analysis in \cite{NS11} claims good agreement with
the \babar data for the $\eta$ TFF using an endpoint-enhanced $\eta$
DA derived from the Nambu--Jona-Lasinio model.
A scheme to describe the $\eta-\eta'$ mixing with two decay parameters
($f_0$ and $f_8$) and two mixing angles ($\theta_0$ and $\theta_8$) was
recently used in~\cite{CIKS12}.
These authors find that their TFF calculation for the two-octet ansatz
is consistent with the bulk of the available data.
At the same time, the $\pi^0$ TFF disagrees with the \babar data for
both the one-octet and the two-octet ansatz.
A more dedicated analysis of the TFFs of the $\eta$-$\eta'$ system
and their mixing properties is given in \cite{KOT12TFF,KOT12}.

\section{Conclusions}
\label{sec:concl}
We presented an analysis that illustrates and discusses the complex
spectrum of challenges encountered in the statistical evaluation
of the various experimental data on the pion-photon TFF and their
interpretation.
Using two different common models (dipole and power-law) to fit the
CELLO, CLEO, \babaR and Belle data, we showed that it is not
possible to predict their trends one from another with acceptable
accuracy.
In particular the rapid growth of the high-$Q^2$ \babar data
cannot be \emph{retrospectively} predicted from the Belle data.
On the other hand, such an auxetic behavior of the TFF as indicated
by the \babar data can hardly be reconciled with the trend of the
Belle data that is compatible with scaling, irrespective of the
fit model used.
In fact, both fit models describe the Belle data with almost the same
statistical precision (Table \ref{tab:table-1}).

From the theoretical side, we studied in this work the effect of a
non-vanishing small virtuality of the quasi-real photon on the TFF
within the LCSR framework.
Though the ensuing suppression of the TFF is rather small for the values
associated with the Belle experiment, this effect is not negligible
--- especially at lower and moderate $Q^2$ values---and
makes it clear that the asymptotic pion DA cannot be considered as a
serious candidate for the description of the data.
In this context let us remark that our predictions in the
intermediate $Q^2$ domain (2-8)~GeV$^2$ also agree in trend with
feasibility studies of the BES-III Collaboration
\cite{U12BESIII} based on a fit to the transition form factor data
of the \babar Collaboration \cite{BaBar09}.
Moreover, the expected high accuracy of this measurement below
5~GeV$^2$\footnote{Marc Unverzagt, private communication.}
will provide a means of excluding pion DAs that yield a scaled TFF with
a steep increase just in this region.
The high accuracy of the data {may also help reducing the
uncertainty in the extraction of the vacuum quark virtuality
$\lambda_{q}^{2}$ that controls the shape of the twist-two pion DA
and also the strength of the twist-four coupling
$\delta_\text{tw-4}^2\approx \lambda_{q}^{2}/2$, see Appendix A in
\cite{BMS02}.

A second theoretical ingredient of our investigation is the inclusion
of the Gegenbauer coefficient $a_6$ into the theoretical scheme to
calculate the pion DA and the pion-photon TFF,
considering its uncertainties as correlated ``noise''.
While a finite photon virtuality influences the TFF predictions at
lower values of $Q^2$ causing suppression, the inclusion of
a third parameter in the representation of the pion DA affects the
result for the TFF mainly in the large-$Q^2$ domain by
increasing the error width of the theoretical band of
predictions.
This suffices to increase even further the compliance of our
predictions with the Belle data (Fig.\ \ref{fig:BMS.3D.FF}).

Indeed, our current investigation in conjunction with our recent
works \cite{BMPS11,BMPS12,BMPS12_LC2012} gives evidence that,
staying within the standard QCD approach based on collinear
factorization, the best overall agreement with all available data
(cf. Fig.\ \ref{fig:pionFF-theor-data}) is provided by BMS-like pion DAs
that represent a compromise between two conflicting urges: to have
enough enhancement in the lower $Q^2$ in order to reach the data from
below, while, on the other hand, to limit that enhancement from above
at higher $Q^2$ so that the scaled TFF saturates and scaling
prevails.
Clearly, such a behavior cannot comply with a power-law which is an
indication that the system is scale-free.
In fact, an auxetic TFF would correspond to a flat-top pion DA
whose main characteristic is that there are no features at some value
of $x$ that makes that particular longitudinal-momentum fraction
stand out.
There are no ``dips'' and ``humps'' anywhere.
This scaling behavior with $x$ lends itself to a pion
interpretation as being a ``pointlike'' particle \cite{RRBGGT10}, in
the sense that it behaves as a unit without revealing its internal
constituents, despite the large momentum $Q^2$ with which it was
probed.

It is worth mentioning that the predictions for the pion TFF
extracted from two independent AdS/QCD approaches
\cite{GR08,BCT11} (see also \cite{BTK2013})
disagree with the high-$Q^2$ trend of the \babar data, while being
in good agreement with the results of the BMS formalism.
A full scale explanation of the antithetic trend of the \babar data
relative to Belle and to the \babar data on the $\eta(\eta')-\gamma$
transition within the confines of QCD cannot be given at present.
New experimental data on the spacelike TFFs of the pseudoscalar mesons
$\pi^0, \eta, \eta'$ could provide a litmus test of the corresponding
DAs of these hadrons shedding also light on the underlying mechanisms
of QCD to create them.

\acknowledgments
A.V.P. is thankful to Prof. Vicente Vento and Prof. Pedro Gonzalez
for the warm hospitality at the University of Valencia, where
part of this investigation was carried out.
The work of A.V.P. was supported in part by the Ministry of Education
and Science of the Russian Federation, project 14.B37.21.0910,
and by HadronPhysics2, Spanish Ministerio de Economia y Competitividad
and EU FEDER under contract FPA2010-21750-C02-01, AIC10-D-000598, and
GVPrometeo2009/129.
We acknowledge support from the Heisenberg--Landau Program under
Grants 2012 and 2013, the Russian Foundation for Fundamental
Research (Grant No.\ 12-02-00613a,\ 11-01-00182a).

\begin{appendix}
\appendix

\section{Form Factors}
\label{sec:FFTs}

We have worked out in Eq.\ (\ref{eq:LCSR-Teylor}) the transition
form factor and its first derivative with respect to $q^2$.
Their explicit forms are
\begin{eqnarray}\nonumber
  \mathcal{F}^{\gamma\pi}\left(Q^2\right)
=&& \!\!\!\!\!
  \frac{\sqrt{2}}{3}f_\pi
  \left[
          \frac{Q^2}{m_{\rho}^2}
          \int_{x_{0}}^{1}
                         \!\! \exp\left(
                                    \frac{m_{\rho}^2-Q^2\bar{x}/x}{M^2}
                              \right) \right.
\\\label{eq:LCSR-FF}
&&\hspace{-18mm}
    \times \left.          \bar{\rho}(Q^2,x) \frac{dx}{x}
         + \int_{0}^{x_0}
                          \!\!\!\bar{\rho}(Q^2,x) \frac{dx}{\bar{x}}
    \right]\!,
\end{eqnarray}
%Eq (A1)
\begin{eqnarray}  \nonumber
  -(\mathcal{F}^{\gamma\pi})'_{q^2=0}\left(Q^2\right)=
  && \!\!\!\!\!
  \frac{\sqrt{2}}{3}f_\pi \!
    \left[
          \frac{Q^2}{m_{\rho}^4}\!
          \int_{x_{0}}^{1}\!
                          \exp\left(
                                    \frac{m_{\rho}^2-Q^2\bar{x}/x}{M^2}
                              \right) \right.
\\ \label{eq:LCSR-derF}
&&\hspace{-18mm}
 \times \left.\bar{\rho}(Q^2,x) \frac{dx}{x}
   + \int_{0}^{x_0} \frac{\bar{\rho}(Q^2,x)}{Q^2}
   \frac{x}{\bar{x}^2}dx  \right] \, .
\end{eqnarray}
%Eq (A2)

With the above expressions one can compute the susceptibility
(``linear response'') $\Delta(Q^2)$ defined in Eq.\ (\ref{eq:delta.Q2}).
The common spectral density $\bar{\rho}$ for the sum of the twist-two
and twist-four contributions is
\begin{eqnarray}
\bar{\rho}(Q^2,x) \nonumber
    &=&
        \sum_{n=0} a_n(Q^2,\mu^2)\bar{\rho}_n(Q^2,x)\!
\\ \label{eq:barrho}
    &&\hspace{10mm}
    + \frac{\delta^2_\text{tw-4}}{Q^2}x\frac{d}{dx}\varphi^{(4)}(x) \ ,
\\ \nonumber
\bar{\rho}_n(Q^2,x)
    &=&
        \bar{\rho}^{(0)}_n(x) + a_s^{1}~\bar{\rho}^{(1)}_n(Q^2,x)
\\ \nonumber &&\hspace{10mm}
   + a_s^{2}~\bar{\rho}^{(2)}_n(Q^2,x)+ \ldots \, .
\end{eqnarray}
%Eq (A3)
The expansion on the l.h.s. of (\ref{eq:barrho}) corresponds
to the expansion of the twist-two pion DA $\varphi_\pi$
over the set of the Gegenbauer polynomials,
$\psi_n(x)\!=\!6x\bar{x}C^{(3/2)}_{n}(x-\bar{x})$,
$\varphi(x,\mu^2)=\sum_{n=0} a_n(\mu^2)\psi_n(x)$,
with $v(n),v^{b}(n)$ being the eigenvalues of the LO
Efremov-Radyushkin-Brodsky-Lepage (ERBL)
equations \cite{ER80,ER80tmf,BL80},
whereas $G_{nl}$ and $b_{n l}$ are calculable triangular
matrices calculated for the first time in \cite{MS09}
and corrected later in \cite{ABOP10}
(see there for more details).
The relevant expressions read
\begin{widetext}
\begin{subequations}
\label{eq:barrho0}
\begin{eqnarray}
&&\bar{\rho}^{(0)}_n(x)=\psi_n(x);~~ \varphi^{(4)}(x)= \frac{80}{3}x^2(1-x)^2 \ ,\label{eq:barrho1-a}  \\
&& \bar{\rho}^{(1)}_n\left(Q^2,\mu^2_{\rm F};x\right)\frac{1}{C_{\rm F}}
=
    \left\{
                    -3\left[1+v^{b}(n)\right]+\frac{\pi^2}{3}
           + 2v(n)
                    \ln\left( \frac{Q^2}{\mu^2_{\rm F}} \right)
           \right\} \psi_n(x)
\label{eq:barrho1-b}  \\
 && +  \left\{2v(n) \ln\left(\frac{\bar{x}}{x} \right)-\ln^2\left(\frac{\bar{x}}{x}\right)
    \right\} \psi_n(x)
     - 2
               \left[  \sum^n_{l=0,2,\ldots}\!\!\!G_{nl}\psi_l(x)
                 +v(n) \left(\sum^n_{l=0,1,\ldots}\!\!\!b_{n l}\psi_l(x)-3\bar{x}\right)\right]
                 \, .
\label{eq:barrho1-c}
\end{eqnarray}
\end{subequations}
\end{widetext}
%Eqs (A4a), (A4b), (A4c))

\section{Hard part of the TFF}
\label{sec:H-amplitude}
To obtain the contributions to the ``hard part'' of the TFF,
$\Ds \int_{0}^{x} \bar{\rho}_n(Q^2,t) \frac{1}{(\bar{t})}dt$,
Eq.\ (\ref{eq:LCSR-FF}), and those to the corresponding
part of the TFF derivative,
$\Ds \int_{0}^{x} \bar{\rho}_n(Q^2,t) \frac{t}{(\bar{t})^2}dt$,
Eq.\ (\ref{eq:LCSR-derF}), we have to integrate these expressions
by inserting for $\bar{\rho}_n$ Eqs.\ (\ref{eq:barrho0}).
Those terms that are proportional to $\psi_n(x)$ only,
Eqs.\ (\ref{eq:barrho1-a}) and (\ref{eq:barrho1-b}), are
obtained by the following closed-form expressions
\begin{widetext}
\begin{eqnarray}
 I_\text{1H}(n,x)&=&\int_0^x \psi_n(t) \frac{1}{(1-t)}~dt\equiv 6\int_0^x C^{(3/2)}_n(2t-1)~tdt=
\frac1{2}\left\{x C^{(1/2)}_{n+1}(2x-1) \right. \nonumber \\
&&\left. - \frac1{2(2n+3)} \left[ C^{(1/2)}_{n+2}(2x-1)-C^{(1/2)}_{n}(2x-1) \right]\right\}; \label{eq:I1H}\\
 J_\text{1H}(n,x)&=&\int_0^x \psi_n(t) \frac{t}{(1-t)^2}~dt\equiv 6\int_0^x C^{(3/2)}_n(2t-1)\frac{t^2}{(1-t)}~dt
=3\left\{(-1)^{n+1}-(n+1)(n+2)\ln(\bar{x}) \right. \nonumber \\
&&\left. -(x+1)C^{(1/2)}_{n+1}(2 x-1) +
\frac{1}{2 (2 n+3)}\left[C^{(1/2)}_{n+2}(2 x-1)-C^{(1/2)}_n(2 x-1)\right]+S(n,0)-S(n,x) \right\},
\label{eq:J1H}
\end{eqnarray}
\end{widetext}
%Eqs (B1) (B2)
where
$$\Ds S(n,x)
=
  \sum_{m=1}^{n} \frac{(n+m+2)!}{m!~(m+1)!~(n-m)!}\frac{(-1)^m}{m}(1-x)^m \, .
$$
The treatment of the first term of Eq.\ (\ref{eq:barrho1-c}) of
$\bar{\rho}^{(1)}_n$ demands some care (see \cite{MS09} for the
origin of this term and further details).
The results for the TFF and its derivative are given by
\begin{widetext}
\begin{eqnarray}
\int_0^x  \left[ 2v(n) \ln\left(\frac{\bar{t}}{t} \right)-\ln^2\left(\frac{\bar{t}}{t}\right)
\right] \frac{\psi_n(t)}{\bar{t}}dt&=&
6\left[ \left(2v(n) \ln\left(\frac{\bar{t}}{t} \right)-\ln^2\left(\frac{\bar{t}}{t}\right)\right)t\right]_{+(x)}
\otimes C^{(3/2)}_{n}(2t-1)+ \nonumber \\
&& 6C^{(3/2)}_{n}(2x-1)\cdot I_\text{2H}(x),  \label{eq:I2Ha} \\
\int_0^x \left[2v(n) \ln\left(\frac{\bar{t}}{t} \right)-\ln^2\left(\frac{\bar{t}}{t}\right) \right]
\frac{\psi_n(t)t}{(\bar{t})^2}dt &=&
6\left[ \left(2v(n) \ln\left(\frac{\bar{t}}{t} \right)-\ln^2\left(\frac{\bar{t}}{t}\right) \right)
\frac{t^2}{\bar{t}}\right]_{+(x)}
\otimes C^{(3/2)}_{n}(2t-1) + \nonumber \\
&& 6C^{(3/2)}_{n}(2x-1)\cdot J_\text{2H}(x) \, ,
\label{eq:JH2a}
\end{eqnarray}
\end{widetext}
%Eqs (B3) (B4)
where
$f(t)_{+(x)}\otimes g(t)= \int_0^xf(t)\left(g(t)-g(x)\right)dt$.

In these equations we have isolated the purely $x$-dependent terms
in two closed-form expressions, which read
\begin{widetext}
\begin{eqnarray}
  I_\text{2H}(x)
  &=& \int_0^x  \left[ 2v(n) \ln\left(\frac{\bar{t}}{t} \right)-\ln^2\left(\frac{\bar{t}}{t}\right)
\right]t dt=   \label{eq:I2Hab}   \\
  &=&  - v(n) \left[\left(1-x^2\right) \ln(\bar{x})+x (x \ln(x)+1)\right]-\left(\text{Li}_2(x)+ x \ln(x)\right)-
          \nonumber   \\
  && \frac{\ln^2(x)}{2}x^2 + \frac{\ln^2(\bar{x})}{2} \left(1-x^2\right) -\bar{x} ((1+x) \ln(x)+1) \ln(\bar{x}) \, ,
           \nonumber\\
J_\text{2H}(x) &=& \int_0^x  \left[ 2v(n) \ln\left(\frac{\bar{t}}{t} \right)-\ln^2\left(\frac{\bar{t}}{t}\right)
\right]\frac{t^2}{\bar{t}}dt= \label{eq:J2Hab}   \\
&=&\!\!2v(n) \left[2 \text{Li}_2(x)+x^2 \ln(x)+\ln(\bar{x}) \left(-x^2-2 x+2 \ln(x)+3\right)+
  x-\ln^2(\bar{x})+2 x \ln(x)\right]+\ln(\bar{x})
            \nonumber \\
  &&\!\!  \left(-2 \text{Li}_2(x)+x^2 (-\ln(x))-x+\ln^2(x)-2 x \ln(x)+3 \ln(x)+\frac{\pi^2}{3}+1\right)+
  3 \text{Li}_2(x)-2 \text{Li}_3(\bar{x})-2 \text{Li}_3(x) +
            \nonumber \\
  &&\!\! 2   \text{Li}_2(x) \ln(x)+\left(\frac{x^2}{2}+x-2 \ln(x)-\frac{3}{2}\right)
  \ln^2(\bar{x})+\frac{1}{2} x^2 \ln^2(x)+\frac{1}{3} \ln^3(\bar{x})+x \ln^2(x)+x \ln(x)+2 \zeta_3 \, ,
  \nonumber
\end{eqnarray}
\end{widetext}
%Eqs (B5) (B6)
recalling that $v(n)$ and $v^{b}(n)$ are the eigenvalues of the
LO ERBL equations.
These two factors, $I_\text{2H}(x)$ and $J_\text{2H}$, prevail over
the corresponding remainders for large $n$.

\section{Moments vs. coefficients}
\label{sec:mom-coef}
The relation between the Gegenbauer coefficients $a_{2n}$ and  the moments
$\langle \xi^{2m}\rangle=\langle(1-2x)^{2m}\rangle,~(m\leq n)$
is given by
\begin{eqnarray}
a_{2n}&=&\frac{2}{3} \frac{(4n+3)}{(2n+1)(2n+2)2^{2n}} \cdot \sum_{m=0}^{n} (-1)^{(n-m)}
\nonumber \\
    && \times \frac{(2n+2m+1)!}{(n+m)! (n-m)! (2m)!} \langle \xi^{2m}\rangle
\end{eqnarray}
%Eq (C1)
with the following first three coefficients
\begin{eqnarray}
a_{2}&=& \frac{7}{12}\left(5 \langle \xi^{2}\rangle -1\right) \, ,    \\
a_{4}&=& \frac{77}{8}\left(\langle \xi^{4}\rangle
        -\frac{2}3 \langle \xi^{2}\rangle +\frac{1}{21}\right) \, ,   \\
a_{6}&=& \frac{5}{64}
    \left(
          429 \langle \xi^{6}\rangle
        - 495 \langle \xi^{4}\rangle
        + 135 \langle \xi^{2}\rangle
        -5  \right) \, .
\end{eqnarray}
%Eq (C2) (C3) (C4)
\end{appendix}
%\vspace{3mm}
%%%%%%%%%%%%%%%%%%%%%%%%%%%%%%%%%%%%%%%%%%%%%%%%%%%%%%%%%%%%%%%%%%%%%%%%%
%% BibTeX Commands %%%%%%%%%%%%%%%%%%%%%%%%%%%%%%%%%%%%%%%%%%%%%%%%%%%%%%
%%%%%%%%%%%%%%%%%%%%%%%%%%%%%%%%%%%%%%%%%%%%%%%%%%%%%%%%%%%%%%%%%%%%%%%%%
%\bibliographystyle{apsrev}
%\bibliography{pion,lambda,nonloc}

\begin{thebibliography}{67}
\expandafter\ifx\csname natexlab\endcsname\relax\def\natexlab#1{#1}\fi
\expandafter\ifx\csname bibnamefont\endcsname\relax
  \def\bibnamefont#1{#1}\fi
\expandafter\ifx\csname bibfnamefont\endcsname\relax
  \def\bibfnamefont#1{#1}\fi
\expandafter\ifx\csname citenamefont\endcsname\relax
  \def\citenamefont#1{#1}\fi
\expandafter\ifx\csname url\endcsname\relax
  \def\url#1{\texttt{#1}}\fi
\expandafter\ifx\csname urlprefix\endcsname\relax\def\urlprefix{URL }\fi
\providecommand{\bibinfo}[2]{#2}
\providecommand{\eprint}[2][]{\url{#2}}

\bibitem[{\citenamefont{Aubert et~al.}(2009)}]{BaBar09}
\bibinfo{author}{\bibfnamefont{B.}~\bibnamefont{Aubert}} \bibnamefont{et~al.}
  (\bibinfo{collaboration}{The BaBar}), \bibinfo{journal}{Phys. Rev.}
  \textbf{\bibinfo{volume}{D80}}, \bibinfo{pages}{052002}
  (\bibinfo{year}{2009}), \eprint{0905.4778}.

\bibitem[{\citenamefont{Mikhailov and Stefanis}(2009)}]{MS09}
\bibinfo{author}{\bibfnamefont{S.~V.} \bibnamefont{Mikhailov}}
  \bibnamefont{and} \bibinfo{author}{\bibfnamefont{N.~G.}
  \bibnamefont{Stefanis}}, \bibinfo{journal}{Nucl. Phys.}
  \textbf{\bibinfo{volume}{B821}}, \bibinfo{pages}{291} (\bibinfo{year}{2009}),
  \eprint{0905.4004}.

\bibitem[{\citenamefont{Brodsky and Lepage}(1989)}]{BL89}
\bibinfo{author}{\bibfnamefont{S.~J.} \bibnamefont{Brodsky}} \bibnamefont{and}
  \bibinfo{author}{\bibfnamefont{G.~P.} \bibnamefont{Lepage}},
  \bibinfo{journal}{Adv. Ser. Direct. High Energy Phys.}
  \textbf{\bibinfo{volume}{5}}, \bibinfo{pages}{93} (\bibinfo{year}{1989}).

\bibitem[{\citenamefont{Bakulev
  et~al.}(2011{\natexlab{a}})\citenamefont{Bakulev, Mikhailov, Pimikov, and
  Stefanis}}]{BMPS11}
\bibinfo{author}{\bibfnamefont{A.~P.} \bibnamefont{Bakulev}},
  \bibinfo{author}{\bibfnamefont{S.~V.} \bibnamefont{Mikhailov}},
  \bibinfo{author}{\bibfnamefont{A.~V.} \bibnamefont{Pimikov}},
  \bibnamefont{and} \bibinfo{author}{\bibfnamefont{N.~G.}
  \bibnamefont{Stefanis}}, \bibinfo{journal}{Phys. Rev.}
  \textbf{\bibinfo{volume}{D84}}, \bibinfo{pages}{034014}
  (\bibinfo{year}{2011}{\natexlab{a}}), \eprint{1105.2753}.

\bibitem[{\citenamefont{Stefanis et~al.}(2012)\citenamefont{Stefanis, Bakulev,
  Mikhailov, and Pimikov}}]{SBMP11}
\bibinfo{author}{\bibfnamefont{N.}~\bibnamefont{Stefanis}},
  \bibinfo{author}{\bibfnamefont{A.~P.} \bibnamefont{Bakulev}},
  \bibinfo{author}{\bibfnamefont{S.}~\bibnamefont{Mikhailov}},
  \bibnamefont{and} \bibinfo{author}{\bibfnamefont{A.}~\bibnamefont{Pimikov}},
  \bibinfo{journal}{Nucl. Phys. Proc. Suppl.}
  \textbf{\bibinfo{volume}{225-227}}, \bibinfo{pages}{146}
  (\bibinfo{year}{2012}), \eprint{1111.7137}.

\bibitem[{\citenamefont{Agaev et~al.}(2011)\citenamefont{Agaev, Braun, Offen,
  and Porkert}}]{ABOP10}
\bibinfo{author}{\bibfnamefont{S.~S.} \bibnamefont{Agaev}},
  \bibinfo{author}{\bibfnamefont{V.~M.} \bibnamefont{Braun}},
  \bibinfo{author}{\bibfnamefont{N.}~\bibnamefont{Offen}}, \bibnamefont{and}
  \bibinfo{author}{\bibfnamefont{F.~A.} \bibnamefont{Porkert}},
  \bibinfo{journal}{Phys. Rev.} \textbf{\bibinfo{volume}{D83}},
  \bibinfo{pages}{054020} (\bibinfo{year}{2011}), \eprint{1012.4671}.

\bibitem[{\citenamefont{Bakulev et~al.}(2001)\citenamefont{Bakulev, Mikhailov,
  and Stefanis}}]{BMS01}
\bibinfo{author}{\bibfnamefont{A.~P.} \bibnamefont{Bakulev}},
  \bibinfo{author}{\bibfnamefont{S.~V.} \bibnamefont{Mikhailov}},
  \bibnamefont{and} \bibinfo{author}{\bibfnamefont{N.~G.}
  \bibnamefont{Stefanis}}, \bibinfo{journal}{Phys. Lett.}
  \textbf{\bibinfo{volume}{B508}}, \bibinfo{pages}{279} (\bibinfo{year}{2001}),
  \eprint[http://arXiv.org/abs]{hep-ph/0103119}.

\bibitem[{\citenamefont{Uehara et~al.}(2012)}]{Belle12}
\bibinfo{author}{\bibfnamefont{S.}~\bibnamefont{Uehara}} \bibnamefont{et~al.}
  (\bibinfo{collaboration}{Belle Collaboration}), \bibinfo{journal}{Phys. Rev.}
  \textbf{\bibinfo{volume}{D86}}, \bibinfo{pages}{092007}
  (\bibinfo{year}{2012}), \eprint{1205.3249}.

\bibitem[{\citenamefont{Behrend et~al.}(1991)}]{CELLO91}
\bibinfo{author}{\bibfnamefont{H.~J.} \bibnamefont{Behrend}}
  \bibnamefont{et~al.} (\bibinfo{collaboration}{CELLO}), \bibinfo{journal}{Z.
  Phys.} \textbf{\bibinfo{volume}{C49}}, \bibinfo{pages}{401}
  (\bibinfo{year}{1991}).

\bibitem[{\citenamefont{Gronberg et~al.}(1998)}]{CLEO98}
\bibinfo{author}{\bibfnamefont{J.}~\bibnamefont{Gronberg}} \bibnamefont{et~al.}
  (\bibinfo{collaboration}{CLEO}), \bibinfo{journal}{Phys. Rev.}
  \textbf{\bibinfo{volume}{D57}}, \bibinfo{pages}{33} (\bibinfo{year}{1998}),
  \eprint{hep-ex/9707031}.

\bibitem[{\citenamefont{Bakulev et~al.}(2012)\citenamefont{Bakulev, Mikhailov,
  Pimikov, and Stefanis}}]{BMPS12}
\bibinfo{author}{\bibfnamefont{A.~P.} \bibnamefont{Bakulev}},
  \bibinfo{author}{\bibfnamefont{S.~V.} \bibnamefont{Mikhailov}},
  \bibinfo{author}{\bibfnamefont{A.~V.} \bibnamefont{Pimikov}},
  \bibnamefont{and} \bibinfo{author}{\bibfnamefont{N.~G.}
  \bibnamefont{Stefanis}}, \bibinfo{journal}{Phys. Rev.}
  \textbf{\bibinfo{volume}{D86}}, \bibinfo{pages}{031501}
  (\bibinfo{year}{2012}), \eprint{1205.3770}.

\bibitem[{\citenamefont{Pimikov et~al.}(2012)\citenamefont{Pimikov, Bakulev,
  Mikhailov, and Stefanis}}]{PBMS12}
\bibinfo{author}{\bibfnamefont{A.}~\bibnamefont{Pimikov}},
  \bibinfo{author}{\bibfnamefont{A.}~\bibnamefont{Bakulev}},
  \bibinfo{author}{\bibfnamefont{S.}~\bibnamefont{Mikhailov}},
  \bibnamefont{and} \bibinfo{author}{\bibfnamefont{N.}~\bibnamefont{Stefanis}},
  \bibinfo{journal}{AIP Conf. Proc.} \textbf{\bibinfo{volume}{1492}},
  \bibinfo{pages}{134} (\bibinfo{year}{2012}), \eprint{1208.4754}.

\bibitem[{\citenamefont{Bakulev et~al.}(2013)\citenamefont{Bakulev, Mikhailov,
  Pimikov, and Stefanis}}]{BMPS12_LC2012}
\bibinfo{author}{\bibfnamefont{A.}~\bibnamefont{Bakulev}},
  \bibinfo{author}{\bibfnamefont{S.}~\bibnamefont{Mikhailov}},
  \bibinfo{author}{\bibfnamefont{A.}~\bibnamefont{Pimikov}}, \bibnamefont{and}
  \bibinfo{author}{\bibfnamefont{N.}~\bibnamefont{Stefanis}},
  \bibinfo{journal}{Acta Phys. Polon. Supp.} \textbf{\bibinfo{volume}{6}},
  \bibinfo{pages}{137} (\bibinfo{year}{2013}), \eprint{1212.0644}.

\bibitem[{\citenamefont{Bakulev
  et~al.}(2004{\natexlab{a}})\citenamefont{Bakulev, Mikhailov, and
  Stefanis}}]{BMS03}
\bibinfo{author}{\bibfnamefont{A.~P.} \bibnamefont{Bakulev}},
  \bibinfo{author}{\bibfnamefont{S.~V.} \bibnamefont{Mikhailov}},
  \bibnamefont{and} \bibinfo{author}{\bibfnamefont{N.~G.}
  \bibnamefont{Stefanis}}, \bibinfo{journal}{Phys. Lett.}
  \textbf{\bibinfo{volume}{B578}}, \bibinfo{pages}{91}
  (\bibinfo{year}{2004}{\natexlab{a}}),
  \eprint[http://arXiv.org/abs]{hep-ph/0303039}.

\bibitem[{\citenamefont{Bakulev et~al.}(2006)\citenamefont{Bakulev, Mikhailov,
  and Stefanis}}]{BMS05lat}
\bibinfo{author}{\bibfnamefont{A.~P.} \bibnamefont{Bakulev}},
  \bibinfo{author}{\bibfnamefont{S.~V.} \bibnamefont{Mikhailov}},
  \bibnamefont{and} \bibinfo{author}{\bibfnamefont{N.~G.}
  \bibnamefont{Stefanis}}, \bibinfo{journal}{Phys. Rev.}
  \textbf{\bibinfo{volume}{D73}}, \bibinfo{pages}{056002}
  (\bibinfo{year}{2006}), \eprint{hep-ph/0512119}.

\bibitem[{\citenamefont{Mikhailov and Radyushkin}(1986)}]{MR86}
\bibinfo{author}{\bibfnamefont{S.~V.} \bibnamefont{Mikhailov}}
  \bibnamefont{and} \bibinfo{author}{\bibfnamefont{A.~V.}
  \bibnamefont{Radyushkin}}, \bibinfo{journal}{JETP Lett.}
  \textbf{\bibinfo{volume}{43}}, \bibinfo{pages}{712} (\bibinfo{year}{1986}).

\bibitem[{\citenamefont{Bakulev and Radyushkin}(1991)}]{BR91}
\bibinfo{author}{\bibfnamefont{A.~P.} \bibnamefont{Bakulev}} \bibnamefont{and}
  \bibinfo{author}{\bibfnamefont{A.~V.} \bibnamefont{Radyushkin}},
  \bibinfo{journal}{Phys. Lett.} \textbf{\bibinfo{volume}{B271}},
  \bibinfo{pages}{223} (\bibinfo{year}{1991}).

\bibitem[{\citenamefont{del Amo~Sanchez et~al.}(2011)}]{BaBar11-BMS}
\bibinfo{author}{\bibfnamefont{P.}~\bibnamefont{del Amo~Sanchez}}
  \bibnamefont{et~al.} (\bibinfo{collaboration}{BABAR Collaboration}),
  \bibinfo{journal}{Phys. Rev.} \textbf{\bibinfo{volume}{D84}},
  \bibinfo{pages}{052001} (\bibinfo{year}{2011}), \eprint{1101.1142}.

\bibitem[{\citenamefont{Feldmann et~al.}(1998)\citenamefont{Feldmann, Kroll,
  and Stech}}]{FKS98}
\bibinfo{author}{\bibfnamefont{T.}~\bibnamefont{Feldmann}},
  \bibinfo{author}{\bibfnamefont{P.}~\bibnamefont{Kroll}}, \bibnamefont{and}
  \bibinfo{author}{\bibfnamefont{B.}~\bibnamefont{Stech}},
  \bibinfo{journal}{Phys. Rev.} \textbf{\bibinfo{volume}{D58}},
  \bibinfo{pages}{114006} (\bibinfo{year}{1998}), \eprint{hep-ph/9802409}.

\bibitem[{\citenamefont{Radyushkin}(2009)}]{Rad09}
\bibinfo{author}{\bibfnamefont{A.~V.} \bibnamefont{Radyushkin}},
  \bibinfo{journal}{Phys. Rev.} \textbf{\bibinfo{volume}{D80}},
  \bibinfo{pages}{094009} (\bibinfo{year}{2009}), \eprint{0906.0323}.

\bibitem[{\citenamefont{Polyakov}(2009)}]{Pol09}
\bibinfo{author}{\bibfnamefont{M.~V.} \bibnamefont{Polyakov}},
  \bibinfo{journal}{JETP Lett.} \textbf{\bibinfo{volume}{90}},
  \bibinfo{pages}{228} (\bibinfo{year}{2009}), \eprint{0906.0538}.

\bibitem[{\citenamefont{Bakulev et~al.}(2003)\citenamefont{Bakulev, Mikhailov,
  and Stefanis}}]{BMS02}
\bibinfo{author}{\bibfnamefont{A.~P.} \bibnamefont{Bakulev}},
  \bibinfo{author}{\bibfnamefont{S.~V.} \bibnamefont{Mikhailov}},
  \bibnamefont{and} \bibinfo{author}{\bibfnamefont{N.~G.}
  \bibnamefont{Stefanis}}, \bibinfo{journal}{Phys. Rev.}
  \textbf{\bibinfo{volume}{D67}}, \bibinfo{pages}{074012}
  (\bibinfo{year}{2003}), \eprint[http://arXiv.org/abs]{hep-ph/0212250}.

\bibitem[{\citenamefont{Bakulev
  et~al.}(2011{\natexlab{b}})\citenamefont{Bakulev, Mikhailov, Pimikov, and
  Stefanis}}]{BMPS_QCD2011}
\bibinfo{author}{\bibfnamefont{A.~P.} \bibnamefont{Bakulev}},
  \bibinfo{author}{\bibfnamefont{S.~V.} \bibnamefont{Mikhailov}},
  \bibinfo{author}{\bibfnamefont{A.~V.} \bibnamefont{Pimikov}},
  \bibnamefont{and} \bibinfo{author}{\bibfnamefont{N.~G.}
  \bibnamefont{Stefanis}}, \bibinfo{journal}{Nucl. Phys. B (Proc. Suppl.)}
  \textbf{\bibinfo{volume}{219--220}}, \bibinfo{pages}{133}
  (\bibinfo{year}{2011}{\natexlab{b}}), \eprint{1108.4344}.

\bibitem[{\citenamefont{Agaev et~al.}(2012)\citenamefont{Agaev, Braun, Offen,
  and Porkert}}]{ABOP12}
\bibinfo{author}{\bibfnamefont{S.}~\bibnamefont{Agaev}},
  \bibinfo{author}{\bibfnamefont{V.}~\bibnamefont{Braun}},
  \bibinfo{author}{\bibfnamefont{N.}~\bibnamefont{Offen}}, \bibnamefont{and}
  \bibinfo{author}{\bibfnamefont{F.}~\bibnamefont{Porkert}},
  \bibinfo{journal}{Phys. Rev.} \textbf{\bibinfo{volume}{D86}},
  \bibinfo{pages}{077504} (\bibinfo{year}{2012}), \eprint{1206.3968}.

\bibitem[{\citenamefont{Khodjamirian}(1999)}]{Kho99}
\bibinfo{author}{\bibfnamefont{A.}~\bibnamefont{Khodjamirian}},
  \bibinfo{journal}{Eur. Phys. J.} \textbf{\bibinfo{volume}{C6}},
  \bibinfo{pages}{477} (\bibinfo{year}{1999}), \eprint{hep-ph/9712451}.

\bibitem[{\citenamefont{Bakulev
  et~al.}(2004{\natexlab{b}})\citenamefont{Bakulev, Mikhailov, and
  Stefanis}}]{BMS04kg}
\bibinfo{author}{\bibfnamefont{A.~P.} \bibnamefont{Bakulev}},
  \bibinfo{author}{\bibfnamefont{S.~V.} \bibnamefont{Mikhailov}},
  \bibnamefont{and} \bibinfo{author}{\bibfnamefont{N.~G.}
  \bibnamefont{Stefanis}}, \bibinfo{journal}{Annalen Phys.}
  \textbf{\bibinfo{volume}{13}}, \bibinfo{pages}{629}
  (\bibinfo{year}{2004}{\natexlab{b}}), \eprint{hep-ph/0410138}.

\bibitem[{\citenamefont{Bakulev and Mikhailov}(2002)}]{BM02}
\bibinfo{author}{\bibfnamefont{A.~P.} \bibnamefont{Bakulev}} \bibnamefont{and}
  \bibinfo{author}{\bibfnamefont{S.~V.} \bibnamefont{Mikhailov}},
  \bibinfo{journal}{Phys. Rev.} \textbf{\bibinfo{volume}{D65}},
  \bibinfo{pages}{114511} (\bibinfo{year}{2002}),
  \eprint[http://arXiv.org/abs]{hep-ph/0203046}.

\bibitem[{\citenamefont{Stefanis et~al.}(1999)\citenamefont{Stefanis, Schroers,
  and Kim}}]{SSK99}
\bibinfo{author}{\bibfnamefont{N.~G.} \bibnamefont{Stefanis}},
  \bibinfo{author}{\bibfnamefont{W.}~\bibnamefont{Schroers}}, \bibnamefont{and}
  \bibinfo{author}{\bibfnamefont{H.-C.} \bibnamefont{Kim}},
  \bibinfo{journal}{Phys. Lett.} \textbf{\bibinfo{volume}{B449}},
  \bibinfo{pages}{299} (\bibinfo{year}{1999}), \eprint{hep-ph/9807298}.

\bibitem[{\citenamefont{Chernyak and Zhitnitsky}(1984)}]{CZ84}
\bibinfo{author}{\bibfnamefont{V.~L.} \bibnamefont{Chernyak}} \bibnamefont{and}
  \bibinfo{author}{\bibfnamefont{A.~R.} \bibnamefont{Zhitnitsky}},
  \bibinfo{journal}{Phys. Rept.} \textbf{\bibinfo{volume}{112}},
  \bibinfo{pages}{173} (\bibinfo{year}{1984}).

\bibitem[{\citenamefont{Meli\'{c} et~al.}(2003)\citenamefont{Meli\'{c},
  M{\"u}ller, and Passek-Kumeri\v{c}ki}}]{MMP02}
\bibinfo{author}{\bibfnamefont{B.}~\bibnamefont{Meli\'{c}}},
  \bibinfo{author}{\bibfnamefont{D.}~\bibnamefont{M{\"u}ller}},
  \bibnamefont{and}
  \bibinfo{author}{\bibfnamefont{K.}~\bibnamefont{Passek-Kumeri\v{c}ki}},
  \bibinfo{journal}{Phys. Rev.} \textbf{\bibinfo{volume}{D68}},
  \bibinfo{pages}{014013} (\bibinfo{year}{2003}), \eprint{hep-ph/0212346}.

\bibitem[{\citenamefont{Czy\.z et~al.}(2012)\citenamefont{Czy\.z, Ivashyn,
  Korchin, and Shekhovtsova}}]{CIKS12}
\bibinfo{author}{\bibfnamefont{H.}~\bibnamefont{Czy\.z}},
  \bibinfo{author}{\bibfnamefont{S.}~\bibnamefont{Ivashyn}},
  \bibinfo{author}{\bibfnamefont{A.}~\bibnamefont{Korchin}}, \bibnamefont{and}
  \bibinfo{author}{\bibfnamefont{O.}~\bibnamefont{Shekhovtsova}},
  \bibinfo{journal}{Phys. Rev.} \textbf{\bibinfo{volume}{D85}},
  \bibinfo{pages}{094010} (\bibinfo{year}{2012}), \eprint{1202.1171}.

\bibitem[{\citenamefont{Bakulev and Mikhailov}(1998)}]{BM98}
\bibinfo{author}{\bibfnamefont{A.~P.} \bibnamefont{Bakulev}} \bibnamefont{and}
  \bibinfo{author}{\bibfnamefont{S.~V.} \bibnamefont{Mikhailov}},
  \bibinfo{journal}{Phys. Lett.} \textbf{\bibinfo{volume}{B436}},
  \bibinfo{pages}{351} (\bibinfo{year}{1998}), \eprint{hep-ph/9803298}.

\bibitem[{\citenamefont{Adler}(1969)}]{Adl69}
\bibinfo{author}{\bibfnamefont{S.~L.} \bibnamefont{Adler}},
  \bibinfo{journal}{Phys. Rev.} \textbf{\bibinfo{volume}{177}},
  \bibinfo{pages}{2426} (\bibinfo{year}{1969}).

\bibitem[{\citenamefont{Bell and Jackiw}(1969)}]{BJ69}
\bibinfo{author}{\bibfnamefont{J.~S.} \bibnamefont{Bell}} \bibnamefont{and}
  \bibinfo{author}{\bibfnamefont{R.}~\bibnamefont{Jackiw}},
  \bibinfo{journal}{Nuovo Cim.} \textbf{\bibinfo{volume}{A60}},
  \bibinfo{pages}{47} (\bibinfo{year}{1969}).

\bibitem[{\citenamefont{Brodsky and Lepage}(1981)}]{BL81-2phot}
\bibinfo{author}{\bibfnamefont{S.~J.} \bibnamefont{Brodsky}} \bibnamefont{and}
  \bibinfo{author}{\bibfnamefont{G.~P.} \bibnamefont{Lepage}},
  \bibinfo{journal}{Phys. Rev.} \textbf{\bibinfo{volume}{D24}},
  \bibinfo{pages}{1808} (\bibinfo{year}{1981}).

\bibitem[{\citenamefont{Ruiz~Arriola and Broniowski}(2002)}]{AriBro-02}
\bibinfo{author}{\bibfnamefont{E.}~\bibnamefont{Ruiz~Arriola}}
  \bibnamefont{and}
  \bibinfo{author}{\bibfnamefont{W.}~\bibnamefont{Broniowski}},
  \bibinfo{journal}{Phys. Rev.} \textbf{\bibinfo{volume}{D66}},
  \bibinfo{pages}{094016} (\bibinfo{year}{2002}), \eprint{hep-ph/0207266}.

\bibitem[{\citenamefont{Ruiz~Arriola and Broniowski}(2003)}]{BroArr03}
\bibinfo{author}{\bibfnamefont{E.}~\bibnamefont{Ruiz~Arriola}}
  \bibnamefont{and}
  \bibinfo{author}{\bibfnamefont{W.}~\bibnamefont{Broniowski}},
  \bibinfo{journal}{Phys. Rev.} \textbf{\bibinfo{volume}{D67}},
  \bibinfo{pages}{074021} (\bibinfo{year}{2003}), \eprint{hep-ph/0301202}.

\bibitem[{\citenamefont{Petrov et~al.}(1999)\citenamefont{Petrov, Polyakov,
  Ruskov, Weiss, and Goeke}}]{PPRWG99}
\bibinfo{author}{\bibfnamefont{V.~Y.} \bibnamefont{Petrov}},
  \bibinfo{author}{\bibfnamefont{M.~V.} \bibnamefont{Polyakov}},
  \bibinfo{author}{\bibfnamefont{R.}~\bibnamefont{Ruskov}},
  \bibinfo{author}{\bibfnamefont{C.}~\bibnamefont{Weiss}}, \bibnamefont{and}
  \bibinfo{author}{\bibfnamefont{K.}~\bibnamefont{Goeke}},
  \bibinfo{journal}{Phys. Rev.} \textbf{\bibinfo{volume}{D59}},
  \bibinfo{pages}{114018} (\bibinfo{year}{1999}), \eprint{hep-ph/9807229}.

\bibitem[{\citenamefont{Anikin et~al.}(2000)\citenamefont{Anikin, Dorokhov, and
  Tomio}}]{ADT99}
\bibinfo{author}{\bibfnamefont{I.}~\bibnamefont{Anikin}},
  \bibinfo{author}{\bibfnamefont{A.}~\bibnamefont{Dorokhov}}, \bibnamefont{and}
  \bibinfo{author}{\bibfnamefont{L.}~\bibnamefont{Tomio}},
  \bibinfo{journal}{Phys. Lett.} \textbf{\bibinfo{volume}{B475}},
  \bibinfo{pages}{361} (\bibinfo{year}{2000}), \eprint{hep-ph/9909368}.

\bibitem[{\citenamefont{Dorokhov}(2010)}]{Dor09}
\bibinfo{author}{\bibfnamefont{A.~E.} \bibnamefont{Dorokhov}},
  \bibinfo{journal}{Phys. Part. Nucl. Lett.} \textbf{\bibinfo{volume}{7}},
  \bibinfo{pages}{229} (\bibinfo{year}{2010}), \eprint{0905.4577}.

\bibitem[{\citenamefont{Dorokhov}(2011)}]{Dor11}
\bibinfo{author}{\bibfnamefont{A.~E.} \bibnamefont{Dorokhov}}
  (\bibinfo{year}{2011}), \bibinfo{note}{1109.3754}, \eprint{1109.3754}.

\bibitem[{\citenamefont{Li and Mishima}(2009)}]{LiMi09}
\bibinfo{author}{\bibfnamefont{H.-n.} \bibnamefont{Li}} \bibnamefont{and}
  \bibinfo{author}{\bibfnamefont{S.}~\bibnamefont{Mishima}},
  \bibinfo{journal}{Phys. Rev.} \textbf{\bibinfo{volume}{D80}},
  \bibinfo{pages}{074024} (\bibinfo{year}{2009}), \eprint{0907.0166}.

\bibitem[{\citenamefont{Arriola and Broniowski}(2010)}]{ArBr10}
\bibinfo{author}{\bibfnamefont{E.~R.} \bibnamefont{Arriola}} \bibnamefont{and}
  \bibinfo{author}{\bibfnamefont{W.}~\bibnamefont{Broniowski}},
  \bibinfo{journal}{Phys. Rev.} \textbf{\bibinfo{volume}{D81}},
  \bibinfo{pages}{094021} (\bibinfo{year}{2010}), \eprint{1004.0837}.

\bibitem[{\citenamefont{Noguera and Vento}(2010)}]{NV10}
\bibinfo{author}{\bibfnamefont{S.}~\bibnamefont{Noguera}} \bibnamefont{and}
  \bibinfo{author}{\bibfnamefont{V.}~\bibnamefont{Vento}},
  \bibinfo{journal}{Eur. Phys. J.} \textbf{\bibinfo{volume}{A46}},
  \bibinfo{pages}{197} (\bibinfo{year}{2010}), \eprint{1001.3075}.

\bibitem[{\citenamefont{Wu and Huang}(2010)}]{WH10}
\bibinfo{author}{\bibfnamefont{X.-G.} \bibnamefont{Wu}} \bibnamefont{and}
  \bibinfo{author}{\bibfnamefont{T.}~\bibnamefont{Huang}},
  \bibinfo{journal}{Phys. Rev.} \textbf{\bibinfo{volume}{D82}},
  \bibinfo{pages}{034024} (\bibinfo{year}{2010}), \eprint{1005.3359}.

\bibitem[{\citenamefont{Kroll}(2011)}]{Kro10sud}
\bibinfo{author}{\bibfnamefont{P.}~\bibnamefont{Kroll}}, \bibinfo{journal}{Eur.
  Phys. J.} \textbf{\bibinfo{volume}{C71}}, \bibinfo{pages}{1623}
  (\bibinfo{year}{2011}), \eprint{1012.3542}.

\bibitem[{\citenamefont{Wu and Huang}(2011)}]{WH11}
\bibinfo{author}{\bibfnamefont{X.-G.} \bibnamefont{Wu}} \bibnamefont{and}
  \bibinfo{author}{\bibfnamefont{T.}~\bibnamefont{Huang}},
  \bibinfo{journal}{Phys. Rev.} \textbf{\bibinfo{volume}{D84}},
  \bibinfo{pages}{074011} (\bibinfo{year}{2011}), \eprint{1106.4365}.

\bibitem[{\citenamefont{Pham and Pham}(2011)}]{PP11}
\bibinfo{author}{\bibfnamefont{T.~N.} \bibnamefont{Pham}} \bibnamefont{and}
  \bibinfo{author}{\bibfnamefont{X.~Y.} \bibnamefont{Pham}},
  \bibinfo{journal}{Int. J. Mod. Phys.} \textbf{\bibinfo{volume}{A26}},
  \bibinfo{pages}{4125} (\bibinfo{year}{2011}), \eprint{1101.3177}.

\bibitem[{\citenamefont{Lih and Geng}(2012)}]{LG12}
\bibinfo{author}{\bibfnamefont{C.-C.} \bibnamefont{Lih}} \bibnamefont{and}
  \bibinfo{author}{\bibfnamefont{C.-Q.} \bibnamefont{Geng}},
  \bibinfo{journal}{Phys. Rev.} \textbf{\bibinfo{volume}{C85}},
  \bibinfo{pages}{018201} (\bibinfo{year}{2012}), \eprint{1201.2220}.

\bibitem[{\citenamefont{Guo and Zhao}(2012)}]{GZ11}
\bibinfo{author}{\bibfnamefont{Z.-k.} \bibnamefont{Guo}} \bibnamefont{and}
  \bibinfo{author}{\bibfnamefont{Q.}~\bibnamefont{Zhao}},
  \bibinfo{journal}{Eur. Phys. J.} \textbf{\bibinfo{volume}{C72}},
  \bibinfo{pages}{1964} (\bibinfo{year}{2012}), \eprint{1108.0241}.

\bibitem[{\citenamefont{Balakireva et~al.}(2012)\citenamefont{Balakireva,
  Lucha, and Melikhov}}]{BLM11}
\bibinfo{author}{\bibfnamefont{I.}~\bibnamefont{Balakireva}},
  \bibinfo{author}{\bibfnamefont{W.}~\bibnamefont{Lucha}}, \bibnamefont{and}
  \bibinfo{author}{\bibfnamefont{D.}~\bibnamefont{Melikhov}},
  \bibinfo{journal}{Phys. Rev.} \textbf{\bibinfo{volume}{D85}},
  \bibinfo{pages}{036006} (\bibinfo{year}{2012}), \eprint{1110.6904}.

\bibitem[{\citenamefont{Klopot et~al.}(2011)\citenamefont{Klopot, Oganesian,
  and Teryaev}}]{KOT11JETP}
\bibinfo{author}{\bibfnamefont{Y.}~\bibnamefont{Klopot}},
  \bibinfo{author}{\bibfnamefont{A.}~\bibnamefont{Oganesian}},
  \bibnamefont{and} \bibinfo{author}{\bibfnamefont{O.}~\bibnamefont{Teryaev}},
  \bibinfo{journal}{JETP Lett.} \textbf{\bibinfo{volume}{94}},
  \bibinfo{pages}{729} (\bibinfo{year}{2011}), \eprint{1110.0474}.

\bibitem[{\citenamefont{Zuo and Huang}(2011)}]{ZH11}
\bibinfo{author}{\bibfnamefont{F.}~\bibnamefont{Zuo}} \bibnamefont{and}
  \bibinfo{author}{\bibfnamefont{T.}~\bibnamefont{Huang}},
  \bibinfo{journal}{Eur. Phys. J.} \textbf{\bibinfo{volume}{C72}},
  \bibinfo{pages}{1813} (\bibinfo{year}{2011}), \eprint{1105.6008}.

\bibitem[{\citenamefont{Feldmann}(2000)}]{Fel00}
\bibinfo{author}{\bibfnamefont{T.}~\bibnamefont{Feldmann}},
  \bibinfo{journal}{Int. J. Mod. Phys.} \textbf{\bibinfo{volume}{A15}},
  \bibinfo{pages}{159} (\bibinfo{year}{2000}), \eprint{hep-ph/9907491}.

\bibitem[{\citenamefont{Noguera and Scopetta}(2012)}]{NS11}
\bibinfo{author}{\bibfnamefont{S.}~\bibnamefont{Noguera}} \bibnamefont{and}
  \bibinfo{author}{\bibfnamefont{S.}~\bibnamefont{Scopetta}},
  \bibinfo{journal}{Phys. Rev.} \textbf{\bibinfo{volume}{D85}},
  \bibinfo{pages}{054004} (\bibinfo{year}{2012}), \eprint{1110.6402}.

\bibitem[{\citenamefont{Klopot et~al.}(2013{\natexlab{a}})\citenamefont{Klopot,
  Oganesian, and Teryaev}}]{KOT12TFF}
\bibinfo{author}{\bibfnamefont{Y.}~\bibnamefont{Klopot}},
  \bibinfo{author}{\bibfnamefont{A.}~\bibnamefont{Oganesian}},
  \bibnamefont{and} \bibinfo{author}{\bibfnamefont{O.}~\bibnamefont{Teryaev}},
  \bibinfo{journal}{Phys. Rev.} \textbf{\bibinfo{volume}{D87}},
  \bibinfo{pages}{036013} (\bibinfo{year}{2013}{\natexlab{a}}),
  \eprint{1211.0874}.

\bibitem[{\citenamefont{Klopot et~al.}(2013{\natexlab{b}})\citenamefont{Klopot,
  Oganesian, and Teryaev}}]{KOT12}
\bibinfo{author}{\bibfnamefont{Y.}~\bibnamefont{Klopot}},
  \bibinfo{author}{\bibfnamefont{A.}~\bibnamefont{Oganesian}},
  \bibnamefont{and} \bibinfo{author}{\bibfnamefont{O.}~\bibnamefont{Teryaev}},
  \bibinfo{journal}{Acta Phys. Polon. Supp.} \textbf{\bibinfo{volume}{6}},
  \bibinfo{pages}{145} (\bibinfo{year}{2013}{\natexlab{b}}),
  \eprint{1212.0459}.

\bibitem[{\citenamefont{Unverzagt}(2012)}]{U12BESIII}
\bibinfo{author}{\bibfnamefont{M.}~\bibnamefont{Unverzagt}},
  \bibinfo{journal}{J. Phys. Conf. Ser.} \textbf{\bibinfo{volume}{349}},
  \bibinfo{pages}{012015} (\bibinfo{year}{2012}).

\bibitem[{\citenamefont{Roberts et~al.}(2010)\citenamefont{Roberts, Roberts,
  Bashir, Gutierrez-Guerrero, and Tandy}}]{RRBGGT10}
\bibinfo{author}{\bibfnamefont{H.~L.~L.} \bibnamefont{Roberts}},
  \bibinfo{author}{\bibfnamefont{C.~D.} \bibnamefont{Roberts}},
  \bibinfo{author}{\bibfnamefont{A.}~\bibnamefont{Bashir}},
  \bibinfo{author}{\bibfnamefont{L.~X.} \bibnamefont{Gutierrez-Guerrero}},
  \bibnamefont{and} \bibinfo{author}{\bibfnamefont{P.~C.} \bibnamefont{Tandy}},
  \bibinfo{journal}{Phys. Rev.} \textbf{\bibinfo{volume}{C82}},
  \bibinfo{pages}{065202} (\bibinfo{year}{2010}), \eprint{1009.0067}.

\bibitem[{\citenamefont{Grigoryan and Radyushkin}(2008)}]{GR08}
\bibinfo{author}{\bibfnamefont{H.~R.} \bibnamefont{Grigoryan}}
  \bibnamefont{and} \bibinfo{author}{\bibfnamefont{A.~V.}
  \bibnamefont{Radyushkin}}, \bibinfo{journal}{Phys. Rev.}
  \textbf{\bibinfo{volume}{D78}}, \bibinfo{pages}{115008}
  (\bibinfo{year}{2008}), \eprint{0808.1243}.

\bibitem[{\citenamefont{Brodsky et~al.}(2011)\citenamefont{Brodsky, Cao, and
  de~Teramond}}]{BCT11}
\bibinfo{author}{\bibfnamefont{S.~J.} \bibnamefont{Brodsky}},
  \bibinfo{author}{\bibfnamefont{F.-G.} \bibnamefont{Cao}}, \bibnamefont{and}
  \bibinfo{author}{\bibfnamefont{G.~F.} \bibnamefont{de~Teramond}},
  \bibinfo{journal}{Phys. Rev.} \textbf{\bibinfo{volume}{D84}},
  \bibinfo{pages}{033001} (\bibinfo{year}{2011}), \eprint{1104.3364}.

\bibitem[{\citenamefont{Brodsky et~al.}(2012)\citenamefont{Brodsky,
  de~Teramond, and Karliner}}]{BTK2013}
\bibinfo{author}{\bibfnamefont{S.~J.} \bibnamefont{Brodsky}},
  \bibinfo{author}{\bibfnamefont{G.}~\bibnamefont{de~Teramond}},
  \bibnamefont{and} \bibinfo{author}{\bibfnamefont{M.}~\bibnamefont{Karliner}},
  \bibinfo{journal}{Ann. Rev. Nucl. Part. Sci.} \textbf{\bibinfo{volume}{62}},
  \bibinfo{pages}{1} (\bibinfo{year}{2012}), \eprint{1302.5684}.

\bibitem[{\citenamefont{Efremov and Radyushkin}(1980{\natexlab{a}})}]{ER80}
\bibinfo{author}{\bibfnamefont{A.~V.} \bibnamefont{Efremov}} \bibnamefont{and}
  \bibinfo{author}{\bibfnamefont{A.~V.} \bibnamefont{Radyushkin}},
  \bibinfo{journal}{Phys. Lett.} \textbf{\bibinfo{volume}{B94}},
  \bibinfo{pages}{245} (\bibinfo{year}{1980}{\natexlab{a}}).

\bibitem[{\citenamefont{Efremov and Radyushkin}(1980{\natexlab{b}})}]{ER80tmf}
\bibinfo{author}{\bibfnamefont{A.~V.} \bibnamefont{Efremov}} \bibnamefont{and}
  \bibinfo{author}{\bibfnamefont{A.~V.} \bibnamefont{Radyushkin}},
  \bibinfo{journal}{Theor. Math. Phys.} \textbf{\bibinfo{volume}{42}},
  \bibinfo{pages}{97} (\bibinfo{year}{1980}{\natexlab{b}}).

\bibitem[{\citenamefont{Lepage and Brodsky}(1980)}]{BL80}
\bibinfo{author}{\bibfnamefont{G.~P.} \bibnamefont{Lepage}} \bibnamefont{and}
  \bibinfo{author}{\bibfnamefont{S.~J.} \bibnamefont{Brodsky}},
  \bibinfo{journal}{Phys. Rev.} \textbf{\bibinfo{volume}{D22}},
  \bibinfo{pages}{2157} (\bibinfo{year}{1980}).

\bibitem[{\citenamefont{Broniowski and Arriola}(2009)}]{BA09}
\bibinfo{author}{\bibfnamefont{W.}~\bibnamefont{Broniowski}} \bibnamefont{and}
  \bibinfo{author}{\bibfnamefont{E.~R.} \bibnamefont{Arriola}}, in
  \emph{\bibinfo{booktitle}{Proceedings of the Mini-Workshop ``Problems in
  Multi-Quark States'', Bled, Slovenia, June 29--July 6, 2009}}, edited by
  \bibinfo{editor}{\bibfnamefont{B.}~\bibnamefont{Golli}},
  \bibinfo{editor}{\bibfnamefont{M.}~\bibnamefont{Rosina}}, \bibnamefont{and}
  \bibinfo{editor}{\bibfnamefont{S.}~\bibnamefont{Sirca}}
  (\bibinfo{publisher}{University of Ljubljana and Jozef Stefan Institute},
  \bibinfo{address}{Ljubljana}, \bibinfo{year}{2009}), pp.
  \bibinfo{pages}{20--27}, \eprint{0910.0869}.

\bibitem[{\citenamefont{Lichard}(2011)}]{Lic10}
\bibinfo{author}{\bibfnamefont{P.}~\bibnamefont{Lichard}},
  \bibinfo{journal}{Phys. Rev.} \textbf{\bibinfo{volume}{D83}},
  \bibinfo{pages}{037503} (\bibinfo{year}{2011}), \eprint{1012.5634}.

\end{thebibliography}
\newcommand{\noopsort}[1]{} \newcommand{\printfirst}[2]{#1}
  \newcommand{\singleletter}[1]{#1} \newcommand{\switchargs}[2]{#2#1}

\end{document}